\documentclass[pdflatex,sn-mathphys]{sn-jnl}

\jyear{2021}%

\theoremstyle{thmstyleone}%
\usepackage{caption}
\usepackage{multirow}
\usepackage{amssymb}
\usepackage{enumerate}
\usepackage{enumitem}
\usepackage{mathtools} 
\usepackage{hyperref}
\usepackage{caption}
\usepackage[T1]{fontenc}
\usepackage{varwidth}

\usepackage{graphicx}
\usepackage{textcomp}
\usepackage{xcolor}
\usepackage{xspace}

\theoremstyle{thmstyletwo}%

\theoremstyle{thmstylethree}%

\newcommand{\ourext}{\emph{post disaster REbuilding plAn ProvIdeR} (REPAIR)\xspace}
\newcommand{\our}{REPAIR\xspace}

\begin{document}

\title[REPAIR] {REPAIR Approach for Social-based City Reconstruction Planning in case of natural disasters}


\author*{\fnm{Ghulam} \sur{Mudassir}*}\email{ghulam.mudassir@buckingham.ac.uk}

\author{\fnm{Antinisca} \sur{{Di} Marco}}\email{antinisca.dimarco@univaq.it}
\equalcont{This author contributed equally to this work.}

\author{\fnm{Giordano} \sur{d'Aloisio}}\email{giordano.daloisio@univaq.it}

\affil{\orgdiv{School of Computing}, \orgname{University of Buckingham,}, \orgaddress{\city{Buckingham}, \postcode{MK181EG}, \country{UK}}}
\affil{\orgdiv{Department of Information Engineering Computer Science and Mathematics }, \orgname{University of L'Aquila}, \orgaddress{\city{L'Aquila}, \postcode{67100}, \country{Italy}}}


\abstract{Natural disasters always have several effects on human lives. It is challenging for governments to tackle these incidents and to rebuild the economic, social and physical infrastructures and facilities with the available resources (mainly budget and time). Governments always define plans and policies according to the law and political strategies that should maximise social benefits. The severity of damage and the vast resources needed to bring life back to normality make such reconstruction a challenge. This article is the extension of our previously published work by conducting comprehensive comparative analysis by integrating additional deep learning models plus random agent which is used as a baseline. Our prior research introduced a decision support system by using the Deep Reinforcement Learning technique for the planning of post-disaster city reconstruction, maximizing the social benefit of the reconstruction process, considering available resources, meeting the needs of the broad community stakeholders (like citizens' social benefits and politicians' priorities) and keeping in consideration city's structural constraints (like dependencies among roads and buildings). The proposed approach, named \emph{post disaster REbuilding plAn ProvIdeR} (REPAIR) is generic. It can determine a set of alternative plans for local administrators who select the ideal one to implement, and it can be applied to areas of any extension. 
We show the application of REPAIR in a real use case, i.e., to the L'Aquila reconstruction process, damaged in 2009 by a major earthquake.}




\maketitle
\section{Introduction} \label{sec:intro}

Disasters like earthquakes are always devastating and destructive. They often result in extensive damage to physical infrastructures like buildings, roads and bridges and the loss of people living in the damaged areas. After disaster events, affected people expect some comprehensive reconstruction strategies to return to their everyday life as soon as possible \cite{zheng2012disaster}. 

Making an effective reconstruction plan always requires an untold amount of time and cost, \cite{li2016di} and it is a critical activity for governments and special offices \cite{mudassir2020social}.  

Describing the recovery process more in detail, the activity occurring after a disaster is called \textit{post-disaster recovery phase} and is defined by Contreras et al.~\cite{Contreras2014} as a \textit{complex multi-dimensional, long-term process involving planning, financing, decision-making, and reconstruction}. Furthermore, according to United Nations Development Program (UNDP), the \textit{post-disaster phase} is divided into four sub-phases: \textit{relief}, \textit{early recovery}, \textit{recovery}, and \textit{development}   ~\cite{undp2009}. According to  \cite{kates1977,alexander2006from} in the \textit{relief} phase, the main aim is to save people's lives by conducting search-and-rescue (SAR) operations; in the \textit {early recovery} phase the main activity is the removal of debris, roads rehabilitation and identification of damaged buildings; while in the \textit{recovery} phase essential services like reconstructions of roads, buildings and social relations of affected people are resumed ~\cite{brown2010}; finally, the \textit{development} phase is focused on strengthening economy, urban resilience and on improving quality of life.  

Information like damage level, causalities, and economic loss is vital in helping stakeholders like disaster management officials, government organizations, and industry representatives in decision-making and post-disaster recovery plans \cite{luis2011visual,zheng2012disaster,zheng2011applying,yang2011hierarchical,li2010ontology}.
However, making recovery plans that effectively tackle disaster situations is always challenging.
Firstly, it is difficult to maintain the balance between formal (i.e., permission from authorities,  the vulnerability of buildings and materials) and informal (i.e., design and development plans, and sustainability) requirements that can ensure the rehabilitation of damaged areas. 
Secondly, defining an effective reconstruction plan that optimises the social benefit by also satisfying constraints related to time and cost is not an easy task. 
Thirdly, public decision-makers have to ensure that the reconstruction plan is \textit{sustainable} from a social and environmental point of view \cite{daloisio_indices, mudassir2020social}.
In addition, the social impact that citizens can experience after successfully applying a recovery plan is usually ignored  during the disaster-recovery phase. 

Due to the complexities mentioned above, the main purpose of this work is to facilitate public decision-makers and citizens by providing an innovative decision model called \ourext that can effectively define and evaluate alternative reconstruction plans by considering all stakeholders' needs.
We consider the \our approach part of the \textit{recovery plan}. Hence, it takes into account the priority of the buildings to reconstruct - including physical features of the whole damaged areas, e.g., physical dependencies (damaged roads or bridges) -and constraints like time and cost for reconstruction. In addition, \our embeds social benefits concerning political strategies. 

The multi-dimensional complexity of the  post-disaster reconstruction phase is tackled by \our through a mathematical formulation of the reconstruction phase (Section \ref{sec:formulation}) and by the adoption of a reinforcement learning technique (i.e., Double Deep Q-learning) to solve such model and define a set of post-disaster reconstruction plans that maximizes the social benefit with respect to the defined constraints (Section:\ref{sec:deeplearning}). 
In the meanwhile, \our is generic and it can be applied to wider areas (various municipalities) and generate alternative plans by changing input parameters (i.e, political priority, considered area and social benefits model).

Hence, by using our approach, we aim to answer the following research questions:

\begin{itemize}
    
\item {\textbf{RQ1:}  Which is the best way to embed the political strategies and political priority into the rebuilding planning model?}
\item {\textbf{RQ2:} How can we model local community needs (namely, social benefits) and embed them into the rebuilding planning model?}
\item {\textbf{RQ3:} How can we model the physical dependencies and embed them into the rebuilding planning model?}
\item \textbf{RQ4:} Which is the most efficient approach that, leveraging the defined rebuilding planning model,  provides alternative rebuilding plans on real case studies?
\item \textbf{RQ5:} How do we validate the proposed post-disaster rebuilding planning approach? 
\item \textbf{RQ6:} Which learning algorithm is the most effective and efficient one for the \our approach? 
\end{itemize}
This article extends our already published paper \cite{9529371} by adding the following new contributions: 

\begin{itemize}
    \item a comparison of different reinforcement learning techniques to find the most effective and efficient one. In particular, we compare the following techniques: Double Deep Q-Learning (DDQN), Q-Learning, SARSA and Deep SARSA;
    \item validation of our approach on a real case study, i.e., the reconstruction of L'Aquila city (Italy) after a major earthquake in 2009 that severely damaged the city. The obtained results demonstrate the soundness of the proposed approach; 
    \item a Python implementation of our approach publicly available for future use cases \cite{ghulam_mudassir_repair}.
\end{itemize}

The paper proceeds as follows:  Section \ref{sec:relatedwork} presents related work, Section \ref{sec:key_concepts} defines the key concepts of the approach, Section \ref{sec:methodology} describes the proposed methodology, in Section \ref{sec:evaluation} we show the evaluation of the proposed approach, in Section \ref{sec:discussion} we discuss our approach with respect to our research questions, Section \ref{sec:future} describes some threats of our approach, and finally Section \ref{sec:conclusion}  concludes the paper presenting future work.

\section{Related Work}\label{sec:relatedwork}

In previous research, only limited studies dealt with post-disaster reconstruction planning issues\cite{phdthesis}. For example, Mfon  et al.~\cite{article} proposed a theoretical framework for quick post-disaster reconstructions considering social, economical, and environmental factors, including providing relief to the affected population. Fan et al.~\cite{Fan2023} proposed decision-support model which integrates graph convolutional neural networks with deep reinforcement learning to repair the road network in post-disaster situation. The framework is pre-trained using a variety of simulated damage scenarios so that, when a hazard occurs, it can rapidly determine near-optimal recovery strategies. Xiao et al.~\cite{xiao2023robotic} proposes a robotic crane equipped with a proximal policy optimization based reinforcement learning algorithm for 3D lift path planning to transport construction materials after earthquakes. Two models were developed and evaluated via simulated load and unload tasks. Bilau et al.~\cite{bilau2015framework} proposed a theoretical model for post-disaster planning; Zhou et al. \cite{5418174} use the 3S\footnote{which is rooted in computer science of Remote Sensing (RS), Global Positioning  System (GPS) and Geographic Information System (GIS)} technique for post-disaster reconstruction planning. An et al. \cite{7730111} proposed an algorithm that combines GIS data (Geographic Information System which is taken by detailed survey results from the Institute of Engineering Mechanics (IEM) and China Earthquake Administration (CEA)) and SAR (Synthetic- Aperture Radar) images to estimate damage assessment after any disaster. Similarly, Doi et al. proposed an approach for reconstruction planning using 2D models and 3D drawings, and Mayo et al. compared the Intelligent Master Planning (IMP) approach with Conventional Master Planning (CMP). In addition, many works proposed formal modelling to optimize different aspects of the reconstruction process. Among those, we cite the following: the model proposed by Hidyat et al.\cite{hidayat2010literature} describing stakeholders' critical role in project management during post-disaster reconstruction (e.g., project financing and designing aspects to start the reconstruction process); the model proposed by Colin et al.\cite{davidson2007truths}  which describes the role of community participation in post-earthquake reconstruction planning; a decision-making framework proposed by Tavakkol et al.~\cite{Tavakkol:2016:EFE:3017611.3017624} for post-disaster reconstruction using the entropy method to prioritize the reconstruction of damaged bridges/ roads and buildings on behalf of available data; finally, a post-disaster recovery model proposed by Ghannad et al.~\cite{ghannad2020multiobjective} which considers social and economic factors of the affected community during reconstruction.

All these works differ from our approach since they do not show any model application that should consider every aspect of post-disaster reconstruction planning. In contrast, we apply our approach to the real use case of L'Aquila city reconstruction in Italy, which was affected by an earthquake in 2009. 

Concerning works that actually propose a model application to optimize the reconstruction process, we can classify them among works that use traditional optimization techniques to define the reconstruction process and works that use an agent-based approach. Among the first category, we cite the multi-criteria model (MCDM) proposed by Opricovic et al. ~\cite{Opricovic2002MulticriteriaPO} to analyze and model post-disaster reconstruction planning. The key feature of this model is to select the best alternative reconstruction plan on behalf of defined input parameters, e.g. like, time, cost, damage level and sustainability. Differently from \our, they have used the fuzzy multi-criteria optimization (FMCO) technique to convert qualitative/not-quantifiable variables. Goujon et al.~\cite{7402039} proposed another multi-criteria decision model for post-disaster reconstruction planning with the help of the Myriad tool \cite{labreuche2005miriad} to define the priority of damaged infrastructure and to evaluate reconstruction projects concerning population needs. Differently from \our, they did not consider concepts like \textit{physical dependencies} and \textit{political priorities} in the reconstruction plan. Qiushan et al. \cite{li2019research} developed a post-disaster framework for reconstructing houses/buildings and implemented the policy in the Dujiangyan central city. According to this framework, multiple entities, like markets, local businesses and social institutions, play an essential role in the economic recovery to create job resources for affected people and quickly reconstruct damaged buildings. The proposed framework elaborates on conventional methods not taking into account cutting-edge technologies like machine learning to handle post-earthquake situations, which are instead used by \our. Vahid et al.\cite{akbari2021online} developed an innovative model using a polynomial-time online algorithm to reconstruct roads in a post-earthquake situation. Authors have validated the performance of the proposed algorithm with others on the Istanbul road network, and results show its performance is much superior to others. Sheykhmousa et al.~\cite{Sheykhmousa2019} proposed a conceptual framework for post-disaster recovery assessment using Land Cover and Land Use Change (LCLUC) information. They have used a support vector machine algorithm for feature extractions and experimental analysis performed on Tacloban city, Philippines dataset which got affected by Typhoon Haiyan in 2013. Xiaoming et al. \cite{zhang2010post} have proposed an optimization model for the post-disaster related situation using Bean Optimization Algorithm (BOA) based on fuzzy preference relation. According to this model, public service buildings get priority in reconstruction and then move on to the rest of the infrastructure. The authors have validated their proposed model in post-earthquake reconstruction in China. Unlike these previous works, \our embeds the Double Deep Q-learning technique to determine the plans to propose to decision-makers.

Concerning the second category of works, Yin et al. \cite{yi2010research} developed an agent-based approach to solve construction management problems in devastating situations after a disaster. The proposed approach can work without any assistance of other stakeholders for construction requirements, provide solutions to public project management problems, and ensure the completion of public projects is successful concerning the time frame in post-disaster reconstruction. Similarly to \cite{yi2010research}, Eid et al. \cite{Eid2018} propose a decision framework using agent-based technology. They have used three agents: a residential agent, an economic agent and a state disaster recovery agent (SDRC). SDRC is used to evaluate the recovery plan and to prioritize the aim after running every simulation through aggregated equations. To the best of our knowledge, these are the only approaches that, like \our, use an agent-based approach to manage reconstruction plans. However, differently from \our their approach does not consider constraints like \textit{physical dependencies} or \textit{political strategies} and does not consider the concept of \textit{social benefit}.

Summarising, differently from all these previous works, our approach proposes a model that embeds reconstruction time and cost constraints, the city's physical dependencies, political strategy and priorities, and social benefits of the local community. In addition, our aim is not to select the best plan but to provide decision-makers with alternative plans that satisfy the posed constraints while guaranteeing high social benefits.

The approach we propose is original since it explicitly considers the following key concepts:
\begin{itemize}
\item \emph{Physical dependencies}  among reconstruction units (like bridge/flyover) that impose ordering in the reconstruction plan.
\item \emph{Social benefits} that, regarding the number of people who will use any unit/building,
describe how much the plan is beneficial for the affected community.
\item \emph{Political strategies} that allow us to model and consider political decisions on the generation of a rebuilding plan.  
\end{itemize}

Finally, to the best of our knowledge, \our is the first to apply the Double Deep Q-Learning approach to find a solution to the posed problem.
\section{Definitions of Key Concepts}\label{sec:key_concepts}

In this Section, we define the key concepts used in our approach in more detail. These are \emph{Physical Dependencies}, \emph{Political Strategies} and \emph{Social Benefits}. 

\begin{figure}[h]
\centering
\includegraphics[width=.7\textwidth]{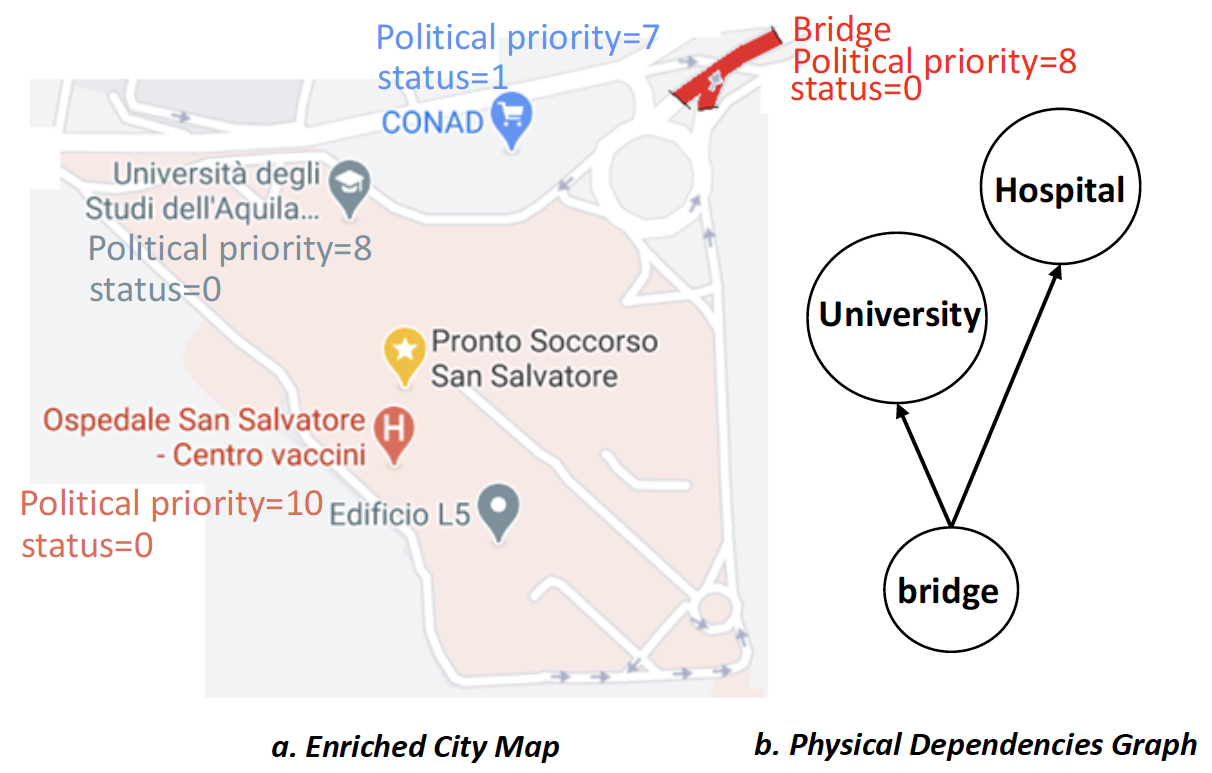}\caption{Physical Dependencies Modeling}\label{fig:phy-depgraph}\end{figure}

\textbf{Physical Dependencies} 
are used to specify a reconstruction priority among damaged reconstruction units (such as buildings, roads, bridges, and others). For example, if any bridge collapses during a natural disaster, it is compulsory to reconstruct it urgently to access other destroyed buildings~\cite{6699055}. Similarly, suppose any natural disaster damages the historical centre of a small town or village. In that case, there could be some physical dependencies among antique houses that are usually very close to each other. That is why the reconstruction order is always compulsory. To formalize all these situations, we introduced the concept of Physical Dependencies in our model. Physical Dependencies are modelled as a directed graph. 
Fig.~\ref{fig:phy-depgraph} shows a simple example that depicts such a scenario. According to the map on the left side of the figure, a bridge is the only possible way to access the hospital, the university and a supermarket (i.e., CONAD). The map contains detailed information, including the status of the city buildings and infrastructure, such as the bridge (status label on the map). Each unit (building/physical dependency) can have two states: a status equal to zero means that the unit has been highly damaged. In contrast, a status equal to one means that the unit has not been highly damaged/reconstructed. According to the map, the bridge, the hospital, and the university buildings have been damaged by a natural disaster (as indicated by \textit{status=0} label). In contrast, the CONAD supermarket does not need to be reconstructed on urgent bases because the damage level is '1' (indicated by \textit{status=1}). The status of the units can be indicated on the map by, for instance, the local municipality of L'Aquila.
In the context depicted in figure \ref{fig:phy-depgraph}, the reconstruction of the hospital or university is not possible because of the damaged bridge, which is the only feasible way to access these buildings. Hence the bridge has a higher priority in the reconstruction. In this case, the bridge is a \emph{physical dependency} since we need to reconstruct it before other damaged buildings (hospital or university). Figure \ref{fig:phy-depgraph} also depicts, on the right, the Physical Dependency Graph. In this graph, each node represents an infrastructure, while the edges represent roads/dependencies connecting the buildings. In this example, the \textit{university} and the \textit{hospital} can be reached only through the \textit{bridge} node. Note that this graph is different from the units graph, which is an undirected graph that represents all the units considered by the model (see Section\ref{sec:dataprocessing}).

\textbf{Political Priority} defines the political strategy to follow. It is related to all reconstruction units and represents each damaged unit's political importance in determining a reconstruction strategy. It is modelled as an attribute whose value is an integer between $[1,10]$. We can represent different political strategies by specifying different political priority values for buildings. For instance, consider this example of political priority: the decision-makers, with the help of policy actors, decide that the reconstruction related to public services must have the same priority as health and educational services. Subsequently, private residences and private and commercial services should be reconstructed. For the implementation of this strategy, higher values are assigned to the political priority attribute of public, health, and education services. In addition, a political priority threshold is assigned to the whole plan (for instance, acceptable plans are the ones whose total political priority is higher than $80\%$) 
Referring to fig.~\ref{fig:phy-depgraph}, here the political strategy is represented by giving a political priority equal to 10 to the hospital, equal to 8 to the bridge and the University of L'Aquila, and equal to $7$ to the CONAD supermarket (in case if it is damaged). 

\textbf{Social Benefits} is defined as ''\emph{the total benefit to society from producing or consuming a good/service. It includes all the private and external benefits of production/consumption.}''\cite{social_ben}. In our proposed research work, Social Benefits is a metric that measures the benefits obtained by the affected local people from a reconstruction plan. In our work, the major aim is to maximize the social benefits during the reconstruction plan definition.  
Formally, the social benefit is defined as the average number of people per day who will get benefits directly and indirectly after the reconstruction plan. For example, when we reconstruct a school, the people who will benefit directly from the reconstruction are primarily the school staff and students. Instead, people that will get indirect benefits are students' parents and relatives. In addition, the obtained benefit can be extended towards hops and services near the school, including people living in the neighbourhood. 

Mathematically, the social benefit of a unit $S(u)$ can be defined as the benefit that the unit $u$ emanates at time $T_u$ as soon as its reconstruction is finished.
Instead, the social benefit of a plan is defined as an aggregated function depending on the order of the reconstruction of the selected units (see Section \ref{sec:formulation} for more details).

\begin{figure}[h]
\centering
\includegraphics[width=9cm,height=4cm]{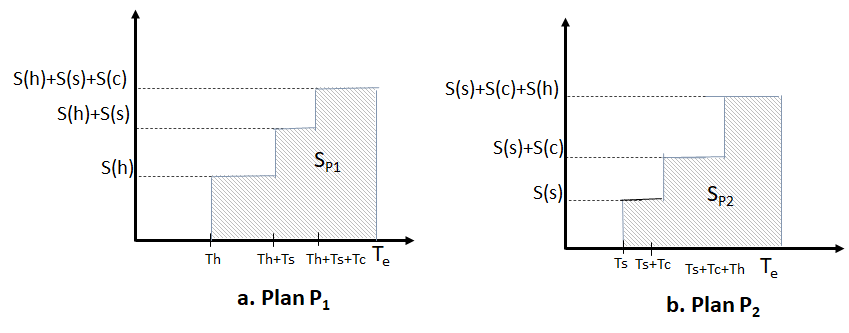}\caption{Social Benefits of Plan $P_1$ and Plan $P_2$}\label{fig:sb-plan}\end{figure}

To better clarify the concept of social benefit, figure \ref{fig:sb-plan} shows the social benefits of two reconstruction plans that contain the same set of reconstruction units (hospital(h), school(s) and cinema(c)). The reconstruction units are rebuilt following a different temporary order. 
In the figure, the x-axis shows the reconstruction time of the units in terms of years. At the start, all of the \emph{recovery} phases (defined in Section\ref{sec:intro}) are zero. $T_e$ is the time we have to complete the reconstruction plan. $T_h$, $T_s$, and $T_c$ represent the time needed to reconstruct the hospital, school and cinema, respectively. For the sake of example, we fix $T_e=6$ years, $T_h=2$ years, $T_s=1.5$ years and $T_c=1$ year. In the figure, the y-axis shows the social benefit of a reconstruction unit.
For the sake of example, we fix $S(h)=2000$, $S(s)=1000$ and $T(c)=600$ people. 

No social benefit is awarded until a unit is reconstructed. In the plan $P_1$, the first unit rebuilt is the 'Hospital', completed in $T_h=2$ years. 
After the completion of 'Hospital', we get $S(h)=2000$ social benefits until the end. Completing the following units (school and cinema) will amplify the total benefit after $T_s=1.5$ and $T_c=1$ years. In this case, the social benefit of the plan is determined by the area dashed in figure \ref{fig:sb-plan}.a that in our example is equal to $11400$ (see Eq. \ref{eqn:max_benefit}). 
In plan $P_2$, we have the same units but in different reconstruction order: at first, the 'School' is reconstructed, then the 'Cinema' and finally the 'Hospital'. In this case, the social benefit of $P_2$ equals $9600$.
Hence, in our example, the most beneficial plan is $P_1$.

\section{Proposed Approach}\label{sec:methodology}
\begin{figure}[th]
\centering
\includegraphics[width=\textwidth]{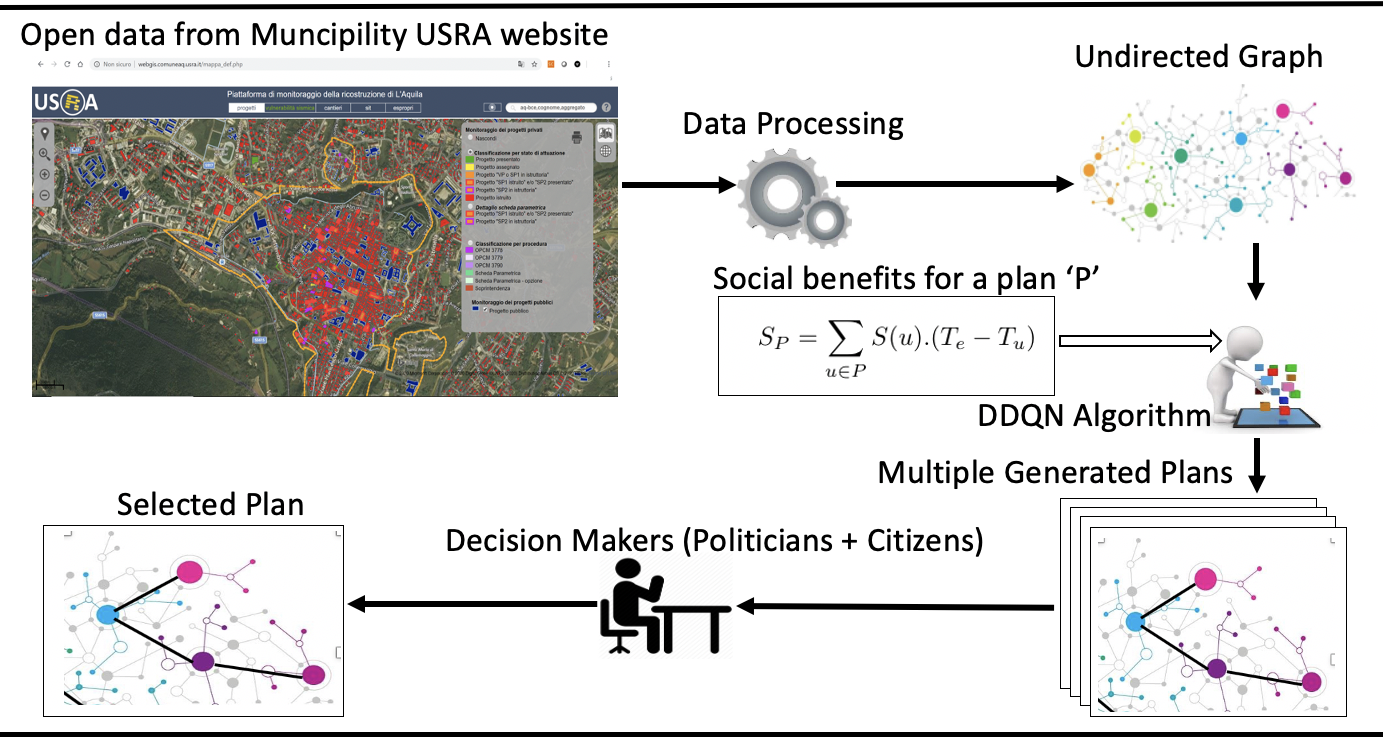}\caption{Proposed Methodology}\label{fig:model}\end{figure}

In this section, we describe our proposed approach (i.e., \our), which is sketched in fig. \ref{fig:model}.
On a high level, it consists of three key steps: 
\emph{i)} formulation of the reconstruction model as an optimization problem; 
\textit{ii)} generation of an undirected graph of the infrastructures from the data obtained by the Web Geographic Information System (WebGIS). Note that this graph is different from the physical dependencies graph described in section \ref{sec:key_concepts});
\emph{iii)} the generation, from the physical dependencies graph and the mathematical reconstruction model, of multiple plans by using deep reinforcement learning (namely, double deep Q-Learning). Eventually, the generated plans are proposed to the decision-makers to select the best one.

In the following sections, we provide details about the three identified phases. The full source code is available on Zenodo \cite{ghulam_mudassir_repair}

\subsection{Model Formulation}\label{sec:formulation}
Our proposed model is based on the infrastructure graph, an undirected graph $G(V,E)$ representing damaged areas. Nodes $v \in V$ indicate the damaged units (such as buildings, roads, bridges, etc.) and edges $e \in E$ represent connections/roads among them. Each node $v$ has a status $s_v \in {0,1} $ that models if $v$ is (still) damaged ($s_v=0$) or not ($s_v=1$).
Here following, we formulate our problem as an optimization model \cite{9529371} that generates a reconstruction plan $P$ by satisfying all the given constraints like time, budget, physical dependencies, etc. and maximizing its social benefits $S_P$. The model is described in detail in Sections \ref{sec:obj_fun} and \ref{sec:model_constr}. It takes as input: the undirected graph of the units, the budget and time limits, and the physical dependencies graph.  
The model returns a set of alternative reconstruction plans that maximize the social benefits of affected communities and respect the given constraints.
Table \ref{tab:Notations} summarises all the variables and attributes that are used in the mathematical model. In the following, we describe the objective function and constraints more in detail.

\begin{table}[ht]
\caption{ 
Mathematical Notations \label{tab:Notations}}
\begin{tabular}{p{0.10\linewidth}p{0.81\linewidth}}\hline

\hline
\textbf{Notation}& \textbf{ Definition}\\
\hline

\hline
$V$	& Set of vertices $v$, that represent  reconstruction units (e.g, building, hospital, bridge).						\\
\hline
$E$	& Set of edge $e$=$(v_1,v_2)$, that represents connections between two reconstruction units, namely $v_1$ and $v_2$.\\
\hline
$s_v$ & $s_v \in \{0,1\}$ represents the status of $v$, if $s_v = 1$ then $v$ is a not damaged or its reconstruction is completed.\\
\hline
$T_v$& the time needed to reconstruct $v$. If $s_v=1$, $T_v$ is zero. \\
  \hline
$c_v$ & Established cost for the reconstruction of v. $s_v=1$, $c_v$ is zero.\\
\hline
$p_v$ & political priority of $v$: $p_v\in\{1, 2 \cdots 10\}$ where 1 represents \emph{Low Priority} and  10 represents \emph{Highest Priority}.\\
\hline
$b_v$ &  Number of people that take direct advantage of $v, b_v \in N$. If $v$ is a residence building, $b_v$ represents the number of people having the residence in $v$. \\
\hline
$S(v)$ &  Number of people that take directly and indirectly advantage after the reconstruction of $v, S(v) \in N$. \\
\hline		
 $d(v_1, v_2)$ & Function returning the distance between $v_1$ and $v_2$ reconstruction units, $v_1$ and $v_2 \in V$x$V$. Such distance is calculated considering the minimum path connecting $v_1$ and $v_2$.\\
  \hline
 $P$ & P represents the generated plan specified as an ordered list of reconstruction units $[v_1, v_2, ..., V_n]$\\
 \hline
  $S_p$ & Social benefits of a plan P\\
 \hline
 $T_e$ & Ending time of the reconstruction plan\\
   \hline
 $G'$ & $G'$=$(V',E')$ is the physical dependencies graph where $V'\subseteq V $ and $E'$ is a set of edge $e'=(v'_1, v'_2)$ representing that $v'_1$ cannot be reconstructed before $v'_2$, as described in Section \ref{sec:key_concepts}. Differently from $G$, $G'$ is a direct graph.	\\  
\hline
\end{tabular}
\end{table}

{\subsubsection{Objective function}\label{sec:obj_fun} 
The reconstruction plan with respect to social benefit $S_P$ is defined as:
\begin{equation}
\label{eqn:max_benefit}
     S_P= \sum_{v \in P}S(v)(T_e - T_v)\\
\end{equation}
where:
\begin{itemize}
\item

$
    S(v) = \begin{cases}
        v \;\;\; &\text{if } v \text{ has not been damaged by the natural disaster}\\
        S_r(v) \;\;\; &\text{otherwise}
    \end{cases}
$

$S_r(v)$ is defined in equation \ref{eq1}. It combines the number of people  taking  benefits directly from the unit $v$ (i.e. $b_v$) with the ones taking benefits indirectly from it. The people taking benefits indirectly are those people living/working in the neighbourhood normalized by distance from the unit $v$. In equation \ref{eq1}, $\alpha$ and $\beta$ are constants that determine the weights of the two social benefit components: $b_v$, and the neighbourhood function.

\begin{equation}
\label{eq1}
\begin{aligned}
    S_r(v)\ &= \left[\alpha \cdot b_v +  \beta\left(\sum\limits_{u \in V \vert s_u = 1}\frac{ S(u)}{ d(u,v)} \right) \right] \\
    &\text{where } \alpha , \beta \in [0,1],  \ \alpha + \beta = 1
\end{aligned}
\end{equation}

\item $(T_e - T_v)$ represents the amount of time needed to reconstruct the unit $v$. In other terms, it represents the time when people start getting the social benefit from the reconstruction of the unit $v$.
\end{itemize}

\subsubsection{Model Constraints}\label{sec:model_constr}
The optimization model proposed in \our has several constraints which must be fulfilled in every reconstruction plan. 

The constraints are the following (in the formulas $P$ is the plan and $v$ are the units):
\begin{itemize}
    \item \textbf{Cost:} it refers to the \emph{Budget} limit to reconstruct units. It is determined by the government's annual financial statements, and it should be less than or equal to the total defined budget set aside for the reconstruction of the territories affected by the natural disaster. Such constraint is formalized as follows:
    \begin{equation}
        \begin{split}
    \sum_{v \in P} C_v \le  Budget
     \end{split}
\end{equation}
    \item \textbf{Plan duration:} the total time required for completion of the reconstruction plan shall be at most equal to the time limit imposed by the government ($T_e$). This constraint is represented as:  
    \begin{equation}
        \sum_{v \in P} T_v \le T_e 
    \end{equation}
   \item \textbf{Political strategy:} as it is described in Section \ref{sec:key_concepts}, our approach also considers political strategies. For this purpose, each unit $v$ is associated with a political priority $P_v$, and each plan must satisfy the following constraint: 
   \begin{equation}
   \frac{\sum_{v \in P} P_v }{\|P\|} \ge Th_p 
\end{equation}
where  $Th_p$ is a threshold to fulfil the required political strategy. In the evaluation reported in Section \ref{sec:evaluation}, we fix $Th_p$ to $8$ for the first cycle of reconstruction plans. This assures that the $80\%$ political strategy must be satisfied in every plan.  

\item \textbf{Physical dependencies:}  every reconstruction plan must comply with the physical dependencies (damage roads/ bridges) represented by the physical dependencies graph $G'=(V', E')$. For example, if any street has a width lower than 3 meters and there are many buildings to be reconstructed, then all the damaged buildings are supposed to be constructed in a specific order that will be decided based on the actual condition of the road. We cannot start from the building in the middle of the street because the next damaged buildings will not be accessible (see also Section \ref{sec:key_concepts}). In general, if a unit $v$ is included in the reconstruction plan, then all the other nodes connected to it in the dependency graph $G'$ must be included too. This constraint is described as: 
\begin{equation}
\forall~v\in P,~\nexists~e \in E' \text{ such that } e=(v,\overline{v}),~s_{\overline{v}}=0~\land~\overline{v}\notin P  
\end{equation}
\end{itemize} 

\subsection{Data Extraction and Processing}\label{sec:dataprocessing}

As mentioned above, \our is based on two data structures: an undirected graph $G$ to represent the damaged areas of a city and a directed graph $G'$ to represent physical dependencies among damaged units. In our approach, we build these two structures starting from shapefiles representing the damaged area to consider \cite{howard2021enriched}. Shapefiles are vectorial data formats used to represent geographical spaces and embed additional information related to a geographical area \cite{harris1998empowerment, mudassir2021toward}.

In particular, in our evaluation, we considered the historical centre of L'Aquila city, which was destroyed by a massive earthquake in 2009. We collected shapefiles from an open online repository\footnote{http://opendata.regione.abruzzo.it/catalog}, and we integrated them with additional shapefiles provided by urban planners. During information extraction, we have also used an adapted version of the work published in~\cite{howard2021enriched} to collect \textit{L'Aquila earthquake} data and \textit{Sulmona} data from shapeflies.
From the collected shapefiles, we are able to extract the graph $G$, the type of units, the number of people using them, and the units' vulnerability index ranging from $0$ to $5$, where $0$ represents the most vulnerable building, and $5$ is the least vulnerable.

However, not all the inputs of our optimization model are available from the shapefiles. In fact, information such as unit status, reconstruction cost and time, and physical dependencies graph, are not derivable from the shapefiles. For this reason, we computed such data in the following way:

\begin{itemize}
    \item the unit status $s_v$ is computed starting from the vulnerability index of the building $v$. In particular, the unit status will be equal to 0 if the vulnerability is between 0 and 3, and will be equal to 1 otherwise;
    \item reconstruction \emph{Cost} and \emph{Time} of a node $v$, is determined by following the approach described by M. Polese et al. in~\cite{POLESE2018139};
    \item \emph{politically priority} is derived from the type of the buildings~\cite{dolce2015building};
    \item the physical dependencies graph $G'$ is derived from the obtained units' status.
\end{itemize}

In the derived graph $G$, nodes representing damaged buildings (i.e., status equal to 0) are shown in red, while fine buildings are shown in green. Instead, edges representing damaged roads are shown with red dotted edges, while not-damaged roads are shown with solid black edges. Algorithm \ref{alg:euclid} shows how the damaged buildings are represented in G.

\begin{algorithm}[tbh]
\renewcommand{\thealgorithm}{1}
\begin{algorithmic}[H]
\caption{Damage Buildings/Roads Identification }
\label{alg:euclid}
 \Require Dataset D, Constructed/Damage Building B, Constructed/Damage Roads R 
        \Ensure  Undirected buildings graph G
         \begin{enumerate}
        \item Extract data from shapefiles
        \item Covert extracted data into .CSV and .XLSX format
         \item Derive the undirected graph $G$
        \end{enumerate}
 \For{v in G}
 \If{v(status) =0} \\
{~~~~~~~~Damage building node will be shown in red color }
 \Else \\
{~~~~~~~~Fine/reconstructed building node will be shown in green }
\EndIf
\EndFor
\For{$R~in~G$}
  read current\;
\If{R(status) =0}\\ 
{~~~~~~~~Damage road will be shown with red dotted edge;}
\Else \\
{~~~~~~~~Fine/reconstructed roads will be shown with dark black edge}
 \EndIf
 \EndFor
\end{algorithmic}
    \end{algorithm}

{\subsection{Reconstruction planning by Double Deep Q-Network (DDQN)} \label{sec:deeplearning} 

Using the obtained values described in section \ref{sec:dataprocessing}, we are able to solve the optimization problem presented in section \ref{sec:formulation}. In particular, in \our we rely on reinforcement learning techniques (Double Deep Q-Network (DDQN)) to solve this problem. We decided to adopt reinforcement learning despite of classical optimization problem solvers because in our literature review, to the best of our knowledge, no one have used reinforcement learning to define post-disaster reconstruction planning. In addition, reinforcement learning agents can be easily adapted to different, and even more complex, situations (see also Section \ref{sec:future}).

DDQN uses convolutional neural networks (CNNs) to approximate action-value non-linear functions called Q-function~\cite{8105597}. In general, in a reinforcement learning problem, we have to provide three main components: \textit{state}, \textit{action}, and \textit{reward} \cite{sutton2018reinforcement}. In \our, they are defined as follows:
\begin{itemize}
    \item  \textbf{State:} is a tuple depicted as \emph{(current location, remaining budget, remaining time)}
    \item \textbf{Action:} represents all possible agent moves in the action space. Making an action corresponds to adding a unit to the reconstruction plan, where each unit is identified by an ID.
    \item \textbf{Reward:} is the social benefit as defined in Eq. \ref{eqn:max_benefit}
\end{itemize}
The reinforcement learning agent uses the reward function to learn the best actions to compute given a specific state. The learning process is performed by updating the Q-Values related to each state-action combination. The Q-Values are updated using the Bellman equation \cite{barron1989bellman}:
\begin{equation}
\label{eqn:qvalue}
Q(s,a; \theta) = S_r(v) + \gamma \max_{a' \in A_v} Q'(s', a'; \theta_i^-)
\end{equation}

where $A_v$ is the set of all actions leading to node $v$.

In the equation, $Q(s,a;\theta)$ is the $Q$ value of taking the action $a$ from state $s$ with weights $\theta$ learned by the neural network. In \our, the state is defined as the combination of four different attributes, namely \emph{(current location and action, remaining budget, remaining time)}, while the action $a$ consists in selecting a \emph{(units/roads ID)}.

$S_r(v)$ is the immediate social reward achieved by taking the optimal action $a$ starting from the current state $s$. Finally, $\gamma$ is the discount factor that trades off the social reward obtained by performing action $a$ with the future reward obtained by performing the action $a'$ on the new state $s'$.  

According to the traditional approach of reinforcement learning \cite{mnih2013playing}, the neural network is trained by minimizing a sequence of loss functions ($L_i(\theta_i)$) that change at each iteration $i$. The loss function in DDQN is the squared difference between Q-target and Q-network.

\begin{equation}
\label{eqn:loss_function}
\begin{split}
L_i(\theta_i) = \mathbb{E} \Bigl[ 
\overbrace{ \bigl( S_r(v) + \gamma \max_{a' \in A_v} Q'(s', a'; \theta_i^-) \bigr) }^{\scriptstyle \text{Q-target}}
- \overbrace{ Q(s, a; \theta_i) }^{\scriptstyle \text{Q-network}} 
\Bigr]^2
\end{split}
\end{equation}

Here, $\theta_i$ is used to compute Q-network, and $\theta_i^-$ is used for Q-target computation. 

In the following, we first describe the implementation of our agent and the base idea behind it. Then we describe the constraints' verification and next we describe the implementation of the entire model.

\subsubsection{Customized Immediate Social Reward Function ($S_r(v)$)
} \label{sec:random_agent}

The main idea of our approach is an agent that selects any damaged unit to reconstruct on behalf of the input parameters (i.e., budget and time ($Te$)) and defined constraints (such as, political priority and physical dependencies)(Eq.\ref{eq1}). As soon as a damaged unit $u$ get reconstructed, the agent searches in the neighbourhood of $u$ to find the next unit `$v$' to reconstruct with the highest social benefits. 

Additionally, if any accessing road/dependency towards `$v$' is damaged, then the dependency/road will be reconstructed first.
The social benefit of the reconstructed dependency (bridge/road) will be added to the following (next reconstructed) unit. 

\subsubsection{Constraints Check}

\begin{algorithm}[ht!]
\caption{Agent Constraints Verification}
\label{alg:algo1}
\begin{algorithmic}[1] 
\Require \textit{Budget} for reconstruction, Time for reconstruction $T_e$
\Ensure All units/buildings $v$ satisfying the constraints
\State Initialize $C_v \gets 0$ and $T_v \gets 0$
\State flag $\gets$ ArrayList$<$Boolean$>$()
\While{$C_v \le \textit{Budget}$ \textbf{and} $T_v \le T_e$}
    \State Keep reconstructing units until \textit{Budget} is exhausted: \newline
    \hspace{1cm}$\textit{Budget} \gets \textit{Budget} - C_v$
    \State Keep reconstructing units until time $T_e$ ends: \newline
    \hspace{1cm}$T_e \gets T_e - T_v$
    \State Political strategies must be satisfied in every reconstruction plan: \newline
    \hspace{1cm}$P_v \ge 8$
    \State Damage to \emph{physical dependencies} is considered in the reconstruction plan
    \State flag.add(\textbf{true})
\EndWhile
\State \textbf{return} flag
\end{algorithmic}
\end{algorithm}

The verification of the constraints is performed each time the agent selects a new node $v$ for the reconstruction and on each reconstruction cycle\footnote{A reconstruction cycle is composed of agent training and a reconstruction plan successful implementation (see Section \ref{sec:planning_impl} for more details). 
}. The formalisation of such verification is depicted in algorithm \ref{alg:algo1}. The function takes as input the total \textit{Budget} and the total time \textit{$T_e$} for the reconstruction and returns the set of nodes satisfying the set of constraints. In particular, the agent continuously adds to the plan damaged units with a high political priority until the budget and time are not consumed completely. 
In addition, the agent also checks if there is any \textit{physical dependency} among the considered damaged units and determines a plan that is compatible with them.
The last constraint to check is the \textit{political priority} ($P_v \ge 8$)~\cite{dolce2015building}. Each building has a specific \emph{political priority} number as shown in Table \ref{tab:buildings priority}. According to this constraint, at least  $80\%$ of the overall political strategy ($Th_p$) should be covered in every cycle (see Table \ref{tab:cycles_priority}). The political strategy threshold is formalised by the following equation:

\begin{equation}
\label{eqn:poltical_priority}
Th_p=(MaxP_v - Training Cycle + 1) \times  \frac{ 80}{100}
\end{equation}

Where $Max{P_v}$ is the defined maximum political priority of `$v$', and $TrainingCycle$ is the round of agent training. The political priority threshold is getting decreased in every cycle (see Table \ref{tab:cycles_priority}) because less beneficial buildings are considered in later cycles. 

\subsubsection{Post-disaster Rebuilding Planning implementation}\label{sec:planning_impl}

\begin{algorithm}
        \begin{algorithmic}[H]
    \caption{Post-disaster Rebuilding Planning}
      \label{alg:ALG2}

\State \textit{\color{blue}{\underline {Training}}}
\begin{itemize}
\item[1.]  Initialise replay memory and Double Deep-Q Network with weights $\theta$
\item[2.]  Find $\bold{B}$: set of unconstructed buildings in G and $\bold{R}$: set of unconstructed roads in G
\end{itemize}

 \For {episode =1, N}
 \State Observe random initial state s 
 \State Pick random node $v$ from B or R
        \While{(size(B)$>$0 $\land$ size(R)$>$0)} 
        \If{random(0,1) $< \epsilon$}
        \State select random action $a$
        \Else 
        \hspace*{5cm}\rlap{\smash{$\left.\begin{array}{@{}c@{}}\\{}\\{}\\{}\\{}\\\end{array}\color{blue}\right\}%
          \color{black}\begin{tabular}{l}{\rotatebox[origin=c]{270}{{$\epsilon$} - greedy action }}\end{tabular}$}}
        \State Select $a$ = ${arg\max_{a\in A_v}}Q{(s,a;{\theta}})$
        \EndIf
        \State Carry out action $a$
        \State Observe reward $S_r(v)$ and new state $s'$
        \State Store experience  $<s, a,S_r(v) ,s'> $ in replay memory
        \State Calculate target for each mini batch
        \State  $
  Set~y_k =
\begin{cases} 
   S_r(v), & \text{for terminal } s' \\  
   S_r(v) + \gamma \max\limits_{a' \in A_v} Q(s', a'; \theta), & \text{otherwise}  
\end{cases}
$
\State\hspace*{41.5em}%
        \rlap{\smash{$\left.\begin{array}{@{}c@{}}\\{}\\{}\\{}\\{}\\{}\\{}\\{}\end{array}\color{blue}\right\}%
          \color{black}\begin{tabular}{l}{\rotatebox[origin=c]{270}{Experience Replay}}\end{tabular}$}}

\State Train the Double Deep-Q Network by performing gradient descent using loss function $(y_k - Q(s, a;\theta))^2$
        \EndWhile
    \EndFor\newline 
------------------------------------------------------------------------------------------------\\
    ~\textit{\color{blue}{\underline{Training verification}}} 
\begin{itemize}
\item[3.]  Verification of trained agent through random agent
\end{itemize}
  \For {episode=1, N}
        \State Run random agent 
        \EndFor 
\begin{itemize}
\item[4.]  Compare random agent results with trained one results
\end{itemize}
 
------------------------------------------------------------------------------------------
~\textit{\color{blue}{\underline{Best alternative reconstruction plans}}}
\begin{itemize}
\item[5.] Apply agent on trained dataset. 
\end{itemize}
     \For {$v \in V$}
        \State Generate reconstruction plan with maximum social benefits $S_p$ (Eq. \ref{eqn:max_benefit}) 
      \EndFor
  \begin{itemize}
         
      \item[6.] Different alternative reconstruction plans are generated.
      \item[7.] Create units sub-lists from plan which can constructed in parallel.
      \item[8.] Update dataset with newly constructed units.
     \item[9.] Update undirected graph G from updated dataset D.
     \item[10.] Save new updated dataset in .XLSX and .CSV file.
     \item[11.] Again agent will be trained on updated data for remaining damage units.
     \end{itemize}
\normalsize
\end{algorithmic}
\end{algorithm}

Algorithm \ref{alg:ALG2} describes the pseudo-code of the rebuilding planner solver by means of Double Deep Q-learning (DDQN). The algorithm is divided into three phases: \textit{training}, \textit{training verification}, and selection of the \textit{best alternative reconstruction plan}.

\textbf{Training.} During the training process, the replay memory (used for the \textit{experience replay} technique \cite{schaul2015prioritized}) and the Deep Q-network model are initialized. Then, the agent first randomly selects a damaged unit $v$ from the set of nodes satisfying the constraints in algorithm \ref{alg:algo1}. Starting from the selected unit $v$, the agent performs an action $a$ using the $\epsilon$-greedy random action selection policy \cite{sutton2018reinforcement}. Then, the agent calculates the social reward $S_r(v)$ and updates the Q-value (see in Eq. \ref{eqn:qvalue}), which is stored in the network. Finally, it moves to the next state $s'$.
However, the approximation of the Q-Value is very unstable and can very easily lead to over fitting~\cite{mnih2013playing}.
To solve this issue, we adopted the \textit{experience replay} technique \cite{schaul2015prioritized}. Following this technique, the agent stores all the experience set $<s, a ,S_r(v),s'>$ during the forward pass. Next, random mini-batches are selected from the replay store to update the set of weights $\theta$. The weights are updated by minimizing the loss function defined in Eq.~\ref{eqn:loss_function}. The loss function is also able to handle imbalanced data \cite{7727770}. The training process continues until all the episodes are performed

\textbf{Training verification.} The verification of the trained agent is performed by comparing its results with the ones of a random agent running for the same number of episodes.

\textbf{Reconstruction plans application.} Once the training is completed, we obtain from the agent a set of \emph{ alternative reconstruction plans} (see in Eq. \ref{eqn:max_benefit}). These plans are obtained from the input dataset. In addition, multiple sub-lists of units which can be constructed in parallel (see Tables \ref{tab:pu1}, \ref{tab:pu2}) are derived from the generated plans.  
After the plans are completed, all the reconstructed units will have their status value changed from 0 to 1. Consequently, a new dataset $D$ is created and a new unit graph $G$ is derived. Again, the agent will be trained on the new dataset for the remaining damaged units, and this process will go on until all units, and roads/dependencies are reconstructed.
\section{Experimental Evaluation}\label{sec:evaluation}
To explain the validity of \our, we applied it to identify a set of reconstruction plans for a specific portion of the historical city centre of L'Aquila (Abruzzo  region, Italy), which was affected by a massive earthquake in 2009.
In our case study, we have considered the damaged area northwest of the crossing of three main streets: \textit{Corso Vittorio Emanuele} (north-south axis of the city), \textit{Corso Principe Umberto}, and \textit{Via San Bernardino}, for a total land size of 246,684.28 m$^2$ as shown in Fig. \ref{fig6:sulmona_city1}. All  the data  was  gathered  from  the  open data portal of the Abruzzo Region (Open GeoData Abruzzo) and  the  project  USRA  (Special  Office  for the Reconstruction of L’Aquila)\footnote{\url{https://bde.comuneaq.usra.it/bdeTrasparente/openData/openDataSet/}}. 
After extracting all the required information from shape files, we converted them into XLSX and CSV formats.
As shown in Fig. \ref{fig6:sulmona_city1}, we represented damaged units and roads as an undirected graph Fig. \ref{fig6:sulmona_graph}. In the graph, red nodes represent damaged units, while green nodes represent fine or reconstructed ones. Similarly, red dotted edges in the graph represent damaged roads and dependencies, while dark black edges show fine/reconstructed ones. The colours of the graph are updated each time a reconstruction plan is completed.
According to the health status of the units, we have found 37 damaged buildings out of 133 and 20 damaged roads out of 150. After the execution of \our, the model complete reconstruction (on behalf of input parameters) in five different cycles of all damaged units/roads by considering maximum social benefits and considering all the defined constraints.  

\begin{figure}[htp]
  \begin{minipage}{0.5\textwidth}
    \centering
   
    \includegraphics[width=1.0\linewidth]{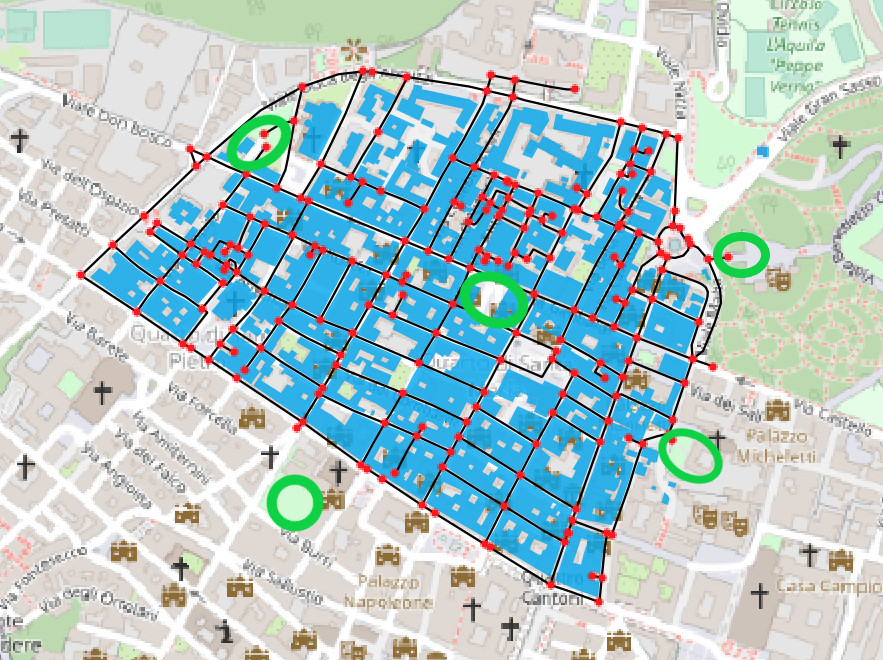}\caption{L'Aquila city map}
    \label{fig6:sulmona_city1}
   
   \end{minipage}\hfill
  \begin{minipage}{0.5\textwidth}
     \centering
     \includegraphics[width=.8\linewidth]{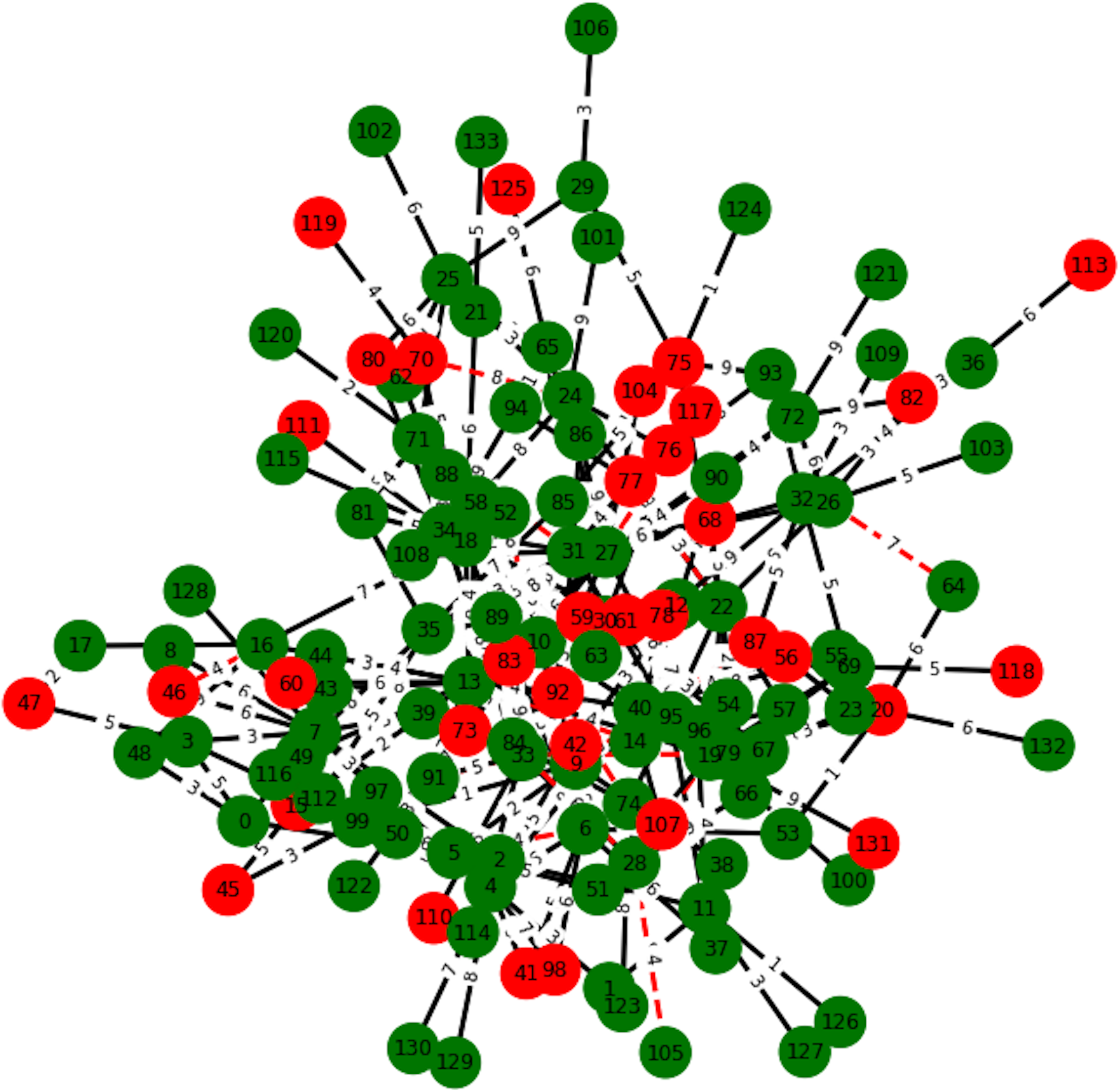} \caption{Damage buildings and roads}
     \label{fig6:sulmona_graph}
   \end{minipage}
\end{figure}

In the following, we first describe DDQN selection and the setup of our experiments; next, we describe the comparison between our solution and a random agent selected as a baseline; and finally, we discuss the results of our experiment relative to our research questions.

\subsection{DDQN and Experimental Setup} \label{sec:experimental_serup}
When applying Double Deep Q-Networks (DDQN) to large-scale post-disaster  reconstruction planning, several practical challenges arise. Chief among them are the significant computational resource requirements and prolonged training durations. So our data is limited that's why all the experiments have been computed by MacBook pro with a quad-core intel core i5 processor and Intel
Iris Plus Graphics 645 graphics. Additionally the performance and efficiency of DDQN is critically dependent on the correct selection of hyperparameter. This is particularly true for computationally intensive tasks like large-scale post-disaster reconstruction, where model efficiency and convergence speed directly affect feasibility. Below, we have mentioned each of the fixed hyperparameter listed in Table \ref{tab:fixed_parameters} .\newline
\textbf{\textit{Adam Optimizer: }}Current setting for Adam Optimizer learning rate = 0.001. Adam is a popular choice due to its adaptive learning rate and fast convergence. However, the learning rate can significantly affect stability. A lower rate may improve stability but slow convergence, while a higher rate risks overshooting. For complex tasks like post-disaster reconstruction planning, where precision is essential and learning rate is beneficial.\newline
\textbf{\textit{Loss Function:}} Current setting for Loss Function is Mean squared error (MSE). MSE is standard for Q-value regression. However, it may suffer from sensitivity to outliers and overestimation bias. .\newline
\textbf{\textit{Q-Learning Function:}} The target calculation uses the target network and mitigates the Q-value overestimation.

Concerning the implementation of our approach, we adopted Python with the Tensorflow library.
The implemented neural network has three fully connected hidden layers and a rectified linear unit activation (relu) function for each. Hidden layers contain 8, 64 and 128 neurons, respectively. Additionally, as it is mentioned before Adam optimizer with a 0.001 learning rate to train the network. The adopted loss function is the mean squared error \cite{allen1971mean}. 
The rebuilding planning agent has been fully trained on 15000 episodes, and the same number of episodes has been used for the evaluation with a random agent. This configuration of the DDQN gave us satisfactory results on the applied dataset. All the hyperparameters of the model are listed in Table \ref{tab:fixed_parameters}.

\begin{table}[ht]
\caption{Fixed Parameters \label{tab:fixed_parameters}}
\begin{tabular}{p{0.303\linewidth}|p{0.63\linewidth}}
\hline

 \textbf{Fixed Parameters} & \textbf{ Value} \\
 \hline
 Optimizer	& Adam optimizer, learning rate = 0.001 						\\
  \hline
  Loss function &	Mean squared error, Eq. \ref{eqn:loss_function}																			\\
  \hline
Q-Learning function & 
$Q(s, a; \theta) = S_r(v) + \gamma \max_{a' \in A_v} Q'(s', a'; \theta_i^-)$ \\
\hline
	
  Batch size &	32																						 \\
 \hline
  Steps before training &	15000																				\\
  \hline
  Maximum memory size &  2000			\\
  \hline			
 	Political Priority &  Minimum=0 , Maximum =10			\\
  \hline
  Exploration strategy & Epsilon greedy policy (Epsilon $\in 10^{-7}, 1$  and self.epsilon\_decay=0.0003.)\\
  \hline
   Reward discount factor &  self.discount\_factor = 0.95
  	\\
  \hline
 Input Parameters &   `Budget' and `Time' ($T_e$)
  	\\
  \hline
\end{tabular}
\end{table}

\subsection{Accuracy Verification and Comparison}

To verify the accuracy of a trained agent in deep reinforcement learning is challenging, and it is still an open field of research \cite{van2017challenges}. Conventional methods to evaluate the accuracy of a reinforcement learning agent are manual verification or the comparison with a random agent \cite{van2017challenges}. In our case, we have adopted the comparison with a random agent because it was a more feasible solution. We ran \our for two reconstruction cycles, and the results are reported in the below figures, where the X-axis shows the number of training episodes (x 100). In contrast, the Y-axis shows the social reward obtained at each iteration. 

In particular, figs.~(\ref{fig6:Plan1-A-Training} ,\ref{fig6:Plan1-B-Training}) represent the Social Reward obtained by the Agent during training for cycle-1 and cycle-2, respectively. In the figures, we report the reward obtained on each episode and the mean reward value. From the figures, it can be seen how, after 15000 episodes, the training curve becomes steeper. It indicates that the Agent is fully trained. 


Figures (\ref{fig6:Plan1-A-Testing}, \ref{fig6:Paln1-B-Testing}) report the comparison between our Agent and a random agent instead. In the figures, the orange curve represents the reward obtained by the random Agent, while the blue curve represents the reward obtained by our Agent. From the figures, it can be seen how our Agent overcomes the reward obtained by a random agent. In particular, the reward obtained by the random Agent ranges between 50 and 75. On the other hand, the trained Agent gets a reward of around 250 for cycle-1 and about 150 for cycle-2


\begin{figure}[ht!]
 \begin{minipage}{0.48\textwidth}
    \centering
    \includegraphics[width=1.1\linewidth]{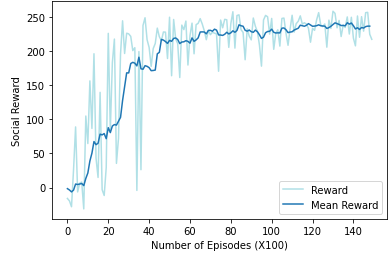}\caption{Cycle-1 Training}
    \label{fig6:Plan1-A-Training}
   
   \end{minipage}\hfill
  \begin{minipage}{0.48\textwidth}
    \centering
    \includegraphics[width=1.1\linewidth]{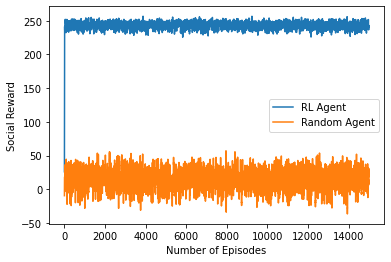}\caption{Cycle-1 Training Verification }
    \label{fig6:Plan1-A-Testing}
   
   \end{minipage}\hfill
 
  \begin{minipage}{0.48\textwidth}
    \centering
    \includegraphics[width=1.1\linewidth]{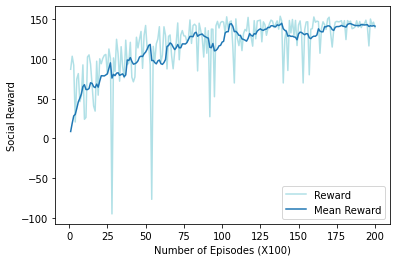}\caption{Cycle-2 Training}
    \label{fig6:Plan1-B-Training}
   \end{minipage}\hfill
  \begin{minipage}{0.48\textwidth}
    \centering
    \includegraphics[width=1.1\linewidth]{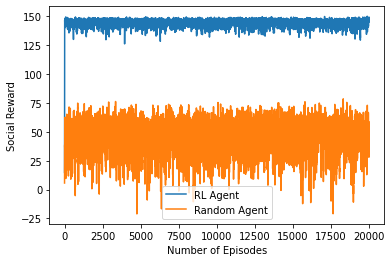}\caption{Cycle-2 Training Verification}
    \label{fig6:Paln1-B-Testing}
   \end{minipage}\hfill
\end{figure}

\section{Discussion}\label{sec:discussion}

In the following, we discuss the impact of our approach concerning our research questions.

{\textbf{RQ1:}} \emph{ Which is the best way to embed the \textbf{political strategies and political priorities} into the rebuilding planning model?}

Modelling political strategies and priorities is one of the key concepts of the rebuilding planning model presented in \our. To model political priorities and strategies, we assigned to each damaged unit a political priority according to their type as shown in Table \ref{tab:buildings priority} ~\cite{dolce2015building}. Each building has a political priority value that ranges between 1 and 10, and different buildings can have the same priority value assigned.
Every proposed reconstruction plan `$S_p$' has a collective political priority ($PP$), which is the sum of all the political priorities of the reconstructed buildings. The collective political priority should meet the thresholds ($Th_P$) defined in Table \ref{tab:cycles_priority}. Each reconstruction cycle has a related threshold which decreases at each cycle. The decrease of $Th_P$ is explained by the fact that in later cycles less beneficial buildings are considered. 
In Table \ref{tab:pp} we show the $PP$ obtained after an actual run of \our for three cycles. It can be seen from the table how the defined thresholds are always satisfied by the solution returned from our model.


  \begin{table}
  \quad \quad \quad \quad
\parbox{.35\textwidth}{
 \caption{Buildings Priority}
   \label{tab:buildings priority}
   \centering
\begin{tabular}{|c|c|}
\hline

 \hline
    \hline
     \textbf{Buildings} & \textbf{Priority}  \\
    \hline

 \hline
    \hline
    Hospitals & 10   \\
    \hline
    Colleges/School & 9 \\
    \hline
    Residential Area & 9\\
    \hline
    Public Points& 8\\
    \hline
    Religious&8\\
     \hline
   Public Buildings& 7\\
    \hline
    Business Centers  & 6  \\
  
    \hline
    Gym Centers & 5 \\
     \hline
    Banquet Halls& 5  \\
      \hline
   Private Buildings & 4  \\
    \hline
  Museums & 3  \\
    \hline
 Bars/Cinemas& 2 \\
     \hline
 Other Places& 1  \\
    \hline

 \hline
    \hline
    
\end{tabular}}
\quad \quad \quad\quad \quad \quad
\parbox{.35\textwidth}{
\centering
\caption{Cycles Political Priority}
   \label{tab:cycles_priority}
\begin{tabular}{|c|c|}
\hline

 \hline
    \hline
\textbf{Cycles} & \textbf{Th\textsubscript{p} (Eq. \ref{eqn:poltical_priority}}) \\
    \hline

 \hline
    \hline
    
    1 & $>=$ 8.0   \\
    \hline
    2 & $>=$ 7.2 \\
    \hline
    3 & $>=$ 6.4\\
    \hline
   4& $>=$ 5.6\\
     \hline
   5& $>=$ 4.8\\
    \hline
    6 & $>=$ 4.0  \\
    \hline
    7 & $>=$ 3.2  \\
    \hline
    8 &  $>=$ 2.4 \\
    \hline
    9 & $>=$ 1.6  \\
    \hline
    10 & $>=$ 0.8  \\
 \hline

 \hline
    \hline
\end{tabular}}
\end{table}

{\textbf{RQ2:} \emph{How can we model local community needs (namely, \textbf{social benefits}) and embed them into the rebuilding planning model?}}

Social benefits '$S_b$' are one of the most important and challenging factor that has to be considered in the reconstruction planning. 
In our approach, the social benefit of every unit to reconstruct is computed considering the number of people `$b_v$' living in that unit plus the number of already reconstructed units `$S_u$' in the neighborhood divided by distance `$d$' (see Eq. \ref{eq1}). 
Further, we computed the total social benefit `$S_p$' of each damaged unit considered in the three reconstruction plans obtained from an actual run of \our (Table \ref{tab:SR}).
Like the thresholds in Table \ref{tab:pp}, also $S_p$ values decreases gradually from C1 to C3. This is explained by the fact that highly beneficial buildings are considered first in the reconstruction process.
\begin{table}
\parbox{.50\textwidth}{
\centering
 \caption{Political Priority}
 \label{tab:pp}
\begin{tabular}{|c|c|c|c|}
\hline
\hline
\hline
 \textbf{Criteria}&  \textbf{ Cycles} & \textbf{PP} & \textbf{Th\textsubscript{P}}\\ \hline\hline
      RQ1& C 1 & 9.4& $>=8.0$ \\ \hline
      RQ1& C 2 & 7.9& $>=7.2$ \\ \hline
      RQ1& C 3 & 6.8& $>=6.4$ \\ \hline
 \hline
 
    \hline
    
\end{tabular}}
\quad
\parbox{.37\textwidth}{
\centering
   \caption{Social Benefits}
   \label{tab:SR}
\begin{tabular}{|c|c|c|}
\hline

 \hline  \hline
    \textbf{Criteria}&\textbf{Cycles} & \textbf{S\textsubscript{P}}  \\   \hline  \hline
     RQ2&C 1  & 3132 \\ \hline
      RQ2&C 2  & 2871\\ \hline
      RQ2& C 3  & 2252\\ \hline
 \hline
 
    \hline
\end{tabular}}
\end{table}

{\textbf{RQ3:}} \emph{How can we model the \textbf{physical city dependencies} and embed them into the rebuilding planning model?}

Physical dependencies are defined as damaged units which must be reconstructed before other units. Examples of physical dependencies are damaged roads or bridges, which allow access to other structures (see the example in section \ref{sec:key_concepts}). In \our, the agent must consider physical dependencies in the definition of a reconstruction plan.

During the definition of a reconstruction plan, the agent also considers roads and bridges as units. If any of them is damaged, then it is included in the reconstruction plan. Damaged physical dependencies will get a higher priority in the reconstruction process than units that are not accessible.  
Physical dependencies are represented by a link between the units' IDs involved in the dependency (e.g, 1-2 means that there is a physical dependency between buildings with IDs 1 and 2).

Concerning our empirical evaluation, Table \ref{tab:cycle-plan1} reports the units involved in the first derived reconstruction. The plan contains 4 physical dependencies (PD), highlighted in boldface. Table \ref{tab:cycle-plan2} instead reports the units considered in the second reconstruction plan. In this case, there are 5 physical dependencies (PD). Note how at least one of the units involved in a physical dependency is not included in the same reconstruction plan. This is explained by the fact that the physical dependency makes at least one of the involved units inaccessible.

In addition to the list of units, \our also returns a list of units that can be reconstructed in parallel. In particular, all the units not involved in a physical dependency can be reconstructed in parallel. In addition, if a damaged physical dependency inhibits access to another physical dependency, the first one will have a higher priority in the reconstruction. 
In Tables \ref{tab:pu1} and \ref{tab:pu2} we report a list of units that can be reconstructed in parallel from cycles 1 and 2, respectively. In particular, the first cycle encompasses 5 lists of parallel reconstruction units, while the second cycle encompasses 6 lists of parallel reconstruction units.

\textbf{RQ4:} \emph{Which is the \textbf{most efficient approach} that, leveraging on the defined rebuilding planning model, provides alternative rebuilding plans on real case studies?}

In \our, we relied on reinforcement learning techniques (i.e. Double Deep Q Networks, DDQN) to solve our optimization model. We preferred this strategy over classical optimization problem-solving techniques because, to the best of our knowledge, this is the first time a reinforcement learning model has been used to solve these kinds of problems. In addition, trained reinforcement learning agents can self-adapt to different and even more contexts (see Section \ref{sec:future}). 

\textbf{RQ5:} \emph{How do we validate the proposed \textbf{post-disaster Rebuilding Planning} approach?}

We proved the capability of \our to define effective (with respect to the obtained social benefit) reconstruction plans by applying it to the actual case study of \textit{historical centre of L'Aquila}. In our experiments, we start with the following input parameters for the first reconstruction cycle: budget \$100,000, and time: 60 months. Once the agent has been trained, it returns two alternative reconstruction plans to satisfy all the defined constraints. Table \ref{tab:cycle-plan1} reports the first plan for the first cycle. It involves 11 buildings with 4 physical dependencies (PD), the total political priority (PP) is 9.4 and the total social benefit `$S_p$' is 3132. Similarly, Table \ref{tab:cycle-plan2} reports the second plan returned for the first cycle of reconstruction. It involves 11 buildings, but this time 5 PD are considered, the total political priority is 9.3, and the total $S_p$ is 3120. The two proposed plans are very similar and have many units in common. It is up to the decision-makers (i.e., municipalities and institutions) the selection the plan to apply.

\begin{table}[!htb]
\parbox{\textwidth}{
\centering
  \caption{Cycle-1:Plan 1}
  \label{tab:cycle-plan1}
\begin{tabular}{c c c c c c c}
\hline
    \multicolumn{7}{|c|}{\textbf{{Budget: \$100,000 Time:60 Months}}}\\\hline
     \textbf{Sr. No} & \textbf{Units ID}& \textbf{Type}&\textbf{Buildings}&\textbf{PD}&\textbf{Total PP}& \textbf{Total S\textsubscript{P}} \\ \hline\hline
     1 & 104&Civil building&&&& \\ \cline{1-3}
     2 &87&Civil building&&&& \\\cline{1-3}
     3 & \textbf{87-9}&P.Dependency&&&& \\ \cline{1-3}
     4 & \textbf{22-77}&P.Dependency&&&& \\ \cline{1-3}
     5 & 77&Civil building&&&& \\ \cline{1-3}
     6 & 82&Civil building&&&& \\ \cline{1-3}
     7 & \textbf{86-65}&P.Dependency&11&4 &9.4&3132\\ \cline{1-3}
     8 & 125&Civil building&&&& \\ \cline{1-3}
     9& 15&Civil building&&&& \\ \cline{1-3}
     10& 41&P.Dependency&&&& \\ \cline{1-3}
      11& 20&Civil building&&&& \\ \cline{1-3}
     12& \textbf{79-33}&P.Dependency&&&& \\ \cline{1-3}
      13& 80&Civil building&&&& \\ \cline{1-3}
     14& 70&Civil building&&&& \\ \cline{1-3}
     15& 83&Civil building&&&& \\ \hline
\hline
\hline
\end{tabular}}
\end{table}

\begin{table}[!htb]
\parbox{\textwidth}{
\centering
 \caption{Cycle-1:Plan 2}
 \label{tab:cycle-plan2}
\begin{tabular}{c c c c c c c}
\hline
    \multicolumn{7}{|c|}{\textbf{{Budget: \$100,000 Time:60 Months}}}\\\hline
     \textbf{Sr. No} & \textbf{Units ID}& \textbf{Type}&\textbf{Buildings}&\textbf{PD}&\textbf{Total PP}& \textbf{Total S\textsubscript{P}} \\  \hline\hline
   1 & 77&Civil building&&&& \\ \cline{1-3}
     2 &\textbf{77-22}&P.Dependency&&&& \\\cline{1-3}
     3 & \textbf{9-87}&P.Dependency&&&& \\ \cline{1-3}
     4 & 87&Civil building&&&& \\ \cline{1-3}
     5 & 82&Civil building&&&& \\ \cline{1-3}
     6 & \textbf{86-65}&P.Dependency&&&& \\ \cline{1-3}
     7 & 125&Civil building&11&5&9.3&3120 \\ \cline{1-3}
     8 & 15&Civil building&&&& \\ \cline{1-3}
     9& 41&Civil building&&&& \\ \cline{1-3}
     10& 20&Civil building&&&& \\ \cline{1-3}
   11& \textbf{79-33}&P.Dependency&&&& \\ \cline{1-3}
     12& 80&Civil building&&&& \\ \cline{1-3}
     13& 70&Civil building&&&& \\ \cline{1-3}
       14 & 83&Civil building&&&& \\ \cline{1-3}
     15& \textbf{16-46}&P.Dependency&&&& \\ \cline{1-3}
     16& 46&Civil building&&&& \\ \hline
\hline
\hline
\end{tabular}}
\end{table}
\begin{table}
\parbox{.50\textwidth}{
\centering
 \caption{Plan 1-Parallel Units }
   \label{tab:pu1}
\begin{tabular}{|c|c|}
\hline
\hline
\hline
     \textbf{Sr.No} & \textbf{Parallel Units} \\ \hline\hline
   1 & [104,87,87-9] \\ \hline
     2 & [22-77] \\ \hline
     3 &[77,82,86-65] \\ \hline
     4 &[125,15,41,20,79-33] \\ \hline
     5 &[80,70,83] \\ \hline
\hline
\hline
    
\end{tabular}}
\quad
\parbox{.45\textwidth}{
   \caption{Plan 2-Parallel Units}
   \label{tab:pu2}
\begin{tabular}{|c|c|}
\hline

    \hline  \hline
\textbf{Sr.No} & \textbf{Parallel Units} \\ \hline\hline
    1 & [77,77-22] \\ \hline
     2 & [9-87] \\ \hline
     3 &[87,82,86-65]\\ \hline
     4 &[125,15,41,20,79-33] \\ \hline
     5 &[80,70,83,16-46] \\ \hline
     6 &[46]\\
 \hline
 \hline
 
    \hline
\end{tabular}}
\end{table}

Finally, Table \ref{tab:cycles_summary} summarizes the output of the other two reconstruction cycles (C2 and C3). For every cycle, we report the total number of units, the number of buildings and the number of PD/roads. In the last two columns, we report the total political priority `PP' and total social benefits `$S_p$' as well. 
\begin{table}[!htbp]
\centering
{
  \caption{Summary Output of C2 and C3 (Budget: \$100,000 Time: 60)}
   \label{tab:cycles_summary}
\begin{tabular}{|c|c|c|c|c|c|c|}
\hline

 \hline
    \hline
     \textbf{ Sr.No} & \textbf{Cycles} & \textbf{Units}&\textbf{ Buildings} & \textbf{ PD/Roads} & \textbf{PP}&\textbf{S\textsubscript{P}} \\ \hline\hline
      
   1 & Cycle 2& 19&12 & 7& 7.9& 2871\\ \hline
     2 & Cycle 3& 21&14& 8& 6.8& 2252 \\ \hline
 \hline
 
    \hline
\end{tabular}}
\end{table} \newline

\textbf{RQ6:} \textit{Which learning algorithm is the most \textbf{effective and efficient} one for \our approach?}

We employed four reinforcement learning algorithms (namely Q-Learning \cite{watkins1992q}, SARSA \cite{zhao2016deep}, Deep SARSA \cite{zhao2016deep}, and Double Deep Q-Networks (DDQN) \cite{8105597}), to identify the most effective and efficient one for our use case. 
In particular, we applied each of these algorithms to solve our optimization model on a small dataset of L’Aquila (including 70 damaged buildings and 27 damaged roads) and measured the social reward obtained by each one. Each algorithm has been implemented in Python using the Keras with Tensorflow framework.  
Fig. \ref{fig:perf_comp} reports the comparison of the obtained social rewards. As shown by the figure, DDQN obtains the highest social reward (around 80). 
\begin{figure}[h]
\centering
\includegraphics[width=.80\textwidth]{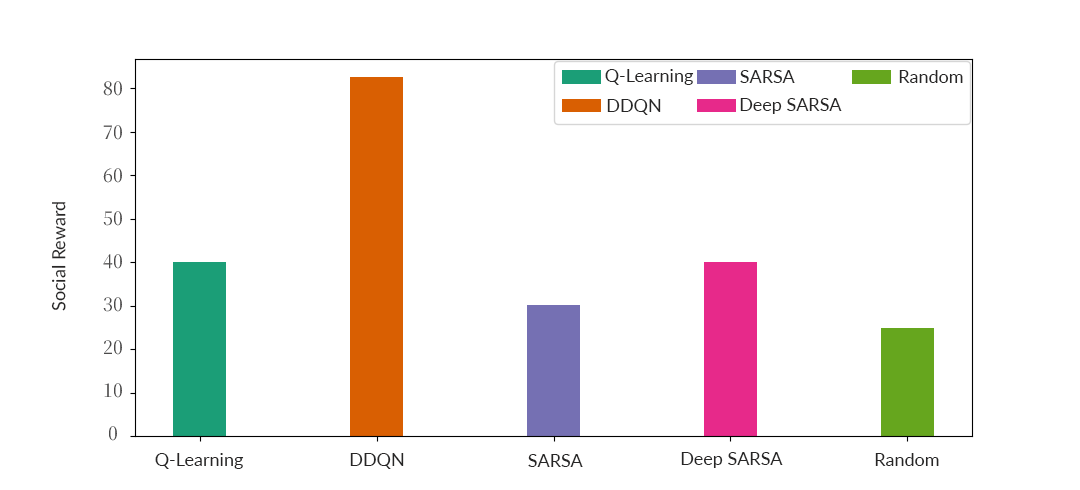}\caption{Performance comparison of RL algorithms}\label{fig:perf_comp}\end{figure}
Table \ref{tab:performance_table} reports a more detailed comparison of these algorithms\footnote {For sake of completeness, in this table we also report the performances of a random agent}. In the table we report: the number of training iterations (which is always 5000), the obtained reward, the number of hours needed for training, and any observed limitation. In particular, we observed that the \textit{Q-Learning} model takes 4 hours to complete 5000 iterations and at the same time memory gets overloaded. Similarly, \textit{SARSA} and \textit{Deep SARSA} training completed in 5.5 and 3.2 hours respectively, and both algorithms presented a high computation complexity. Finally, the \textit{Random Agent} took 3 hours but returned a very low reward compared to the other methods. DDQN instead took 2.5 hours to compute 5000 iterations and obtained the highest reward among all the other methods, i.e., 80. 

\begin{table}[!htbp]
\centering
{
  \caption{Performance comparison table}
   \label{tab:performance_table}
\begin{tabular}{|c|c|c|c|c|c|c|}
\hline

 \hline
    \hline
     \textbf{ Sr.No} & \textbf{Algorithms} &\textbf{ Iterations}& \textbf{Reward}&\textbf{Time(hr)} &   \textbf{Limitations} \\ \hline\hline
     1 &Q-Learning&5000&40&4& Memory overload\\ \hline
    2 & DDQN&5000&80&2.5&Not found\\ \hline
     3 & SARSA & 5000& 38& 5.5&Comput. Complexity \\ \hline
    4 & Deep SARSA & 5000 & 42& 3.2&Comput. Complexity\\ \hline
     5 & Random Agent & 5000 & 25& 3&Low Reward\\ \hline
 \hline
 
    \hline
\end{tabular}}
\end{table}
\section{Threat to Validity}\label{sec:future}

In this section, we describe some limitations of the \our approach. 
 \begin{itemize}
\item\textbf{External Validity: Scalability of our Approach.} We have successfully evaluated our approach to the case study of the L'Aquila earthquake. However, we know that post-disaster situations can be very different from each other and even more complex. To cope with this issue, we relied on reinforcement learning techniques (i.e., DDQN) to solve our proposed optimization model. The adopted approach is generic and can easily be adapted to more complex environments. In particular, a trained agent can adapt to a more complex environment with limited training, relying on the knowledge already acquired.


\item\textbf{External Validity: Generality of our Approach.} Our approach is related to the particular domain of post-disaster management. However, we believe the proposed model can also be adapted to similar management problems where attributes like time, cost and social benefits are considered.

\item\textbf{Internal Validity: Incorrect Forecast.} 
The main internal validity threat of \our is incorrect information about damaged areas that can lead to wrong reconstruction plans. Due to data uncertainty, post-disaster reconstructions can be started in the least affected areas. This will negate the main idea of the proposed framework. We are aware of this issue and have observed it on two different real datasets during our evaluation process.


\end{itemize}

\section{Conclusion and Future work} \label{sec:conclusion}
In this paper, we have proposed a comprehensive decision-support system called \textit{post-disaster Rebuilding Plan Provider (\our)}. This system uses Double Deep Q-learning Network (DDQN) to define a reconstruction plan for damaged units. 

In particular, we first presented the considered reconstruction problem modelled as an optimization model considering variables like time, cost, physical dependencies, social benefits of the affected community and political priority of units. Next, we described the DDQN algorithm used to solve this model. The trained agent returns a set of alternative reconstruction plans which satisfies all the constraints and maximizes the social benefits. We have successfully validated our approach by applying it to the real case study of the \textit{L'Aquila} earthquake. 


In future works, we will proceed in the following directions: 
\begin{itemize}
\item we will consider service access quality and street walk-ability considering more specific attributes than the number of residents living in that area; 
\item we will focus on improving the computational complexity of the reconstruction plans' definition; 
\item we will perform quantitative research to assess damage level accurately by using cutting-edge technologies. This would be helpful in estimating the damaged infrastructure and to evaluate the reconstruction budget accurately.    
\end{itemize}

\vspace{1em}
\small

\noindent \textbf{Acknowledgements -} This work is partially funded by Territori Aperti (a project funded by Fondo Territori Lavoro e Conoscenza CGIL, CSIL and UIL) and by SoBigData-PlusPlus   H2020-INFRAIA-2019-1 EU project, contract number 871042. The open data used in the evaluation comes from \emph{opendata.regione.abruzzo.it}.\newline
--------------------------------------------------------------------------------------------------------------\\

\noindent \textbf{Statements of Declaration:} \newline \\
\textit{\textbf{Ethics approval and consent to participate:}} \\\\
This study did not involve any human participants, human data, or human tissue. Therefore, ethics approval and consent to participate were not required. \\\\
\textit{\textbf{Consent for publication:}} \\\\
Not consent is applicable for publication. 
\\\\
\textit{\textbf{Availability of data and material:}} \\\\
The datasets generated and analyzed during the current study are available at this link \cite{ghulam_mudassir_repair}. 
\\\\
\textit{\textbf{Conflict of interests:}} \\\\
The authors declare that they have no conflict of interests. 
\\\\
\textit{\textbf{Funding:}} \\\\
This work has been partially supported by Territori Aperti (a project funded by Fondo Territori, Lavoro e Conoscenza CGIL CISL UIL) and by European Union — NextGenerationEU — NationalRecovery and Resilience Plan (Piano Nazionale di Ripresa e Resilienza, PNRR) — Project: “So-BigData.it — Strengthening the Italian RI for Social Mining and Big Data Analytics” — Prot. IR0000013 — Avviso n. 3264 del 28/12/2021. 
\\\\
\textit{\textbf{Authors Contribution:}} 
\begin{itemize}
\item \textbf{Ghulam Mudassir:} Methodology, Formal Analysis, Investigation, Software, Data Curation, Writing - original Draft, Writing - review Editing, Visualization
\item \textbf{Antinisca Di Marco:} Supervision, Writing - review editing
\item \textbf{Giordano d’Aloisio:} Writing - review editing
\end{itemize}
All authors have read and approved the final manuscript 
. 
\\\\
\normalsize
\bibliography{bibliography.bib}


\begin{thebibliography}{52}
\ifx \bisbn   \undefined \def \bisbn  #1{ISBN #1}\fi
\ifx \binits  \undefined \def \binits#1{#1}\fi
\ifx \bauthor  \undefined \def \bauthor#1{#1}\fi
\ifx \batitle  \undefined \def \batitle#1{#1}\fi
\ifx \bjtitle  \undefined \def \bjtitle#1{#1}\fi
\ifx \bvolume  \undefined \def \bvolume#1{\textbf{#1}}\fi
\ifx \byear  \undefined \def \byear#1{#1}\fi
\ifx \bissue  \undefined \def \bissue#1{#1}\fi
\ifx \bfpage  \undefined \def \bfpage#1{#1}\fi
\ifx \blpage  \undefined \def \blpage #1{#1}\fi
\ifx \burl  \undefined \def \burl#1{\textsf{#1}}\fi
\ifx \doiurl  \undefined \def \doiurl#1{\url{https://doi.org/#1}}\fi
\ifx \betal  \undefined \def \betal{\textit{et al.}}\fi
\ifx \binstitute  \undefined \def \binstitute#1{#1}\fi
\ifx \binstitutionaled  \undefined \def \binstitutionaled#1{#1}\fi
\ifx \bctitle  \undefined \def \bctitle#1{#1}\fi
\ifx \beditor  \undefined \def \beditor#1{#1}\fi
\ifx \bpublisher  \undefined \def \bpublisher#1{#1}\fi
\ifx \bbtitle  \undefined \def \bbtitle#1{#1}\fi
\ifx \bedition  \undefined \def \bedition#1{#1}\fi
\ifx \bseriesno  \undefined \def \bseriesno#1{#1}\fi
\ifx \blocation  \undefined \def \blocation#1{#1}\fi
\ifx \bsertitle  \undefined \def \bsertitle#1{#1}\fi
\ifx \bsnm \undefined \def \bsnm#1{#1}\fi
\ifx \bsuffix \undefined \def \bsuffix#1{#1}\fi
\ifx \bparticle \undefined \def \bparticle#1{#1}\fi
\ifx \barticle \undefined \def \barticle#1{#1}\fi
\bibcommenthead
\ifx \bconfdate \undefined \def \bconfdate #1{#1}\fi
\ifx \botherref \undefined \def \botherref #1{#1}\fi
\ifx \url \undefined \def \url#1{\textsf{#1}}\fi
\ifx \bchapter \undefined \def \bchapter#1{#1}\fi
\ifx \bbook \undefined \def \bbook#1{#1}\fi
\ifx \bcomment \undefined \def \bcomment#1{#1}\fi
\ifx \oauthor \undefined \def \oauthor#1{#1}\fi
\ifx \citeauthoryear \undefined \def \citeauthoryear#1{#1}\fi
\ifx \endbibitem  \undefined \def \endbibitem {}\fi
\ifx \bconflocation  \undefined \def \bconflocation#1{#1}\fi
\ifx \arxivurl  \undefined \def \arxivurl#1{\textsf{#1}}\fi
\csname PreBibitemsHook\endcsname

\bibitem{zheng2012disaster}
\begin{bchapter}
\bauthor{\bsnm{Zheng}, \binits{L.}},
\bauthor{\bsnm{Shen}, \binits{C.}},
\bauthor{\bsnm{Tang}, \binits{L.}},
\bauthor{\bsnm{Zeng}, \binits{C.}},
\bauthor{\bsnm{Li}, \binits{T.}},
\bauthor{\bsnm{Luis}, \binits{S.}},
\bauthor{\bsnm{Chen}, \binits{S.-C.}},
\bauthor{\bsnm{Navlakha}, \binits{J.K.}}:
\bctitle{Disaster sitrep-a vertical search engine and information analysis tool in disaster management domain}.
In: \bbtitle{2012 IEEE 13th International Conference on Information Reuse \& Integration (IRI)},
pp. \bfpage{457}--\blpage{465}
(\byear{2012}).
\bcomment{IEEE}
\end{bchapter}
\endbibitem

\bibitem{li2016di}
\begin{bchapter}
\bauthor{\bsnm{Li}, \binits{T.}},
\bauthor{\bsnm{Zhou}, \binits{W.}},
\bauthor{\bsnm{Zeng}, \binits{C.}},
\bauthor{\bsnm{Wang}, \binits{Q.}},
\bauthor{\bsnm{Zhou}, \binits{Q.}},
\bauthor{\bsnm{Wang}, \binits{D.}},
\bauthor{\bsnm{Xu}, \binits{J.}},
\bauthor{\bsnm{Huang}, \binits{Y.}},
\bauthor{\bsnm{Wang}, \binits{W.}},
\bauthor{\bsnm{Zhang}, \binits{M.}}, \betal:
\bctitle{Di-dap: an efficient disaster information delivery and analysis platform in disaster management}.
In: \bbtitle{Proceedings of the 25th ACM International on Conference on Information and Knowledge Management},
pp. \bfpage{1593}--\blpage{1602}
(\byear{2016})
\end{bchapter}
\endbibitem

\bibitem{mudassir2020social}
\begin{bchapter}
\bauthor{\bsnm{Mudassir}, \binits{G.}}:
\bctitle{Social-based physical reconstruction planning in case of natural disaster: A machine learning approach}.
In: \bbtitle{International Conference on Research Challenges in Information Science},
pp. \bfpage{604}--\blpage{612}
(\byear{2020}).
\bcomment{Springer}
\end{bchapter}
\endbibitem

\bibitem{Contreras2014}
\begin{barticle}
\bauthor{\bsnm{Contreras}, \binits{D.}},
\bauthor{\bsnm{Blaschke}, \binits{T.}},
\bauthor{\bsnm{Kienberger}, \binits{S.}},
\bauthor{\bsnm{Zeil}, \binits{P.}}:
\batitle{Myths and realities about the recovery of l×³aquila after the earthquake}.
\bjtitle{International Journal of Disaster Risk Reduction}
\bvolume{8},
\bfpage{125}--\blpage{142}
(\byear{2014}).
\doiurl{10.1016/j.ijdrr.2014.02.001}
\end{barticle}
\endbibitem

\bibitem{undp2009}
\begin{botherref}
\oauthor{\bsnm{Programme}, \binits{U.D.}}:
UNDP Policy on Early Recovery.
(2008).
\url{https://www.undp.org/content/dam/undp/library/crisis\%20prevention/undp-cpr-policy-brief-early-recovery-2008-08-22.pdf}
\end{botherref}
\endbibitem

\bibitem{kates1977}
\begin{botherref}
\oauthor{\bsnm{Kates}, \binits{R.}},
\oauthor{\bsnm{Pijawka}, \binits{D.}},
\oauthor{\bsnm{Haas}, \binits{J.}},
\oauthor{\bsnm{Kates}, \binits{R.}},
\oauthor{\bsnm{Bowden}, \binits{M.}}:
Reconstruction following disaster.
Haas, JE, Kates, RW, Bowden, MJ, Eds
(1977)
\end{botherref}
\endbibitem

\bibitem{alexander2006from}
\begin{bchapter}
\bauthor{\bsnm{Alexander}, \binits{D.}}:
\bctitle{From rubble to monument revisited: Modernised perspectives on recovery from disaster}.
In: \bbtitle{Post-disaster Reconstruction: Meeting Stakeholder Interests}.
\bpublisher{Firenze University Press},
\blocation{Florence, Italy}
(\byear{2006})
\end{bchapter}
\endbibitem

\bibitem{brown2010}
\begin{botherref}
\oauthor{\bsnm{Brown}, \binits{D.}},
\oauthor{\bsnm{Platt}, \binits{S.}},
\oauthor{\bsnm{Bevington}, \binits{J.}}:
Disaster recovery indicators: Guidelines for monitoring and evaluation.
Cambridge University Centre for Risk in the Built Environment
(2010)
\end{botherref}
\endbibitem

\bibitem{luis2011visual}
\begin{bchapter}
\bauthor{\bsnm{Luis}, \binits{S.}},
\bauthor{\bsnm{Fleites}, \binits{F.C.}},
\bauthor{\bsnm{Yang}, \binits{Y.}},
\bauthor{\bsnm{Ha}, \binits{H.-Y.}},
\bauthor{\bsnm{Chen}, \binits{S.-C.}}:
\bctitle{A visual analytics multimedia mobile system for emergency response}.
In: \bbtitle{2011 IEEE International Symposium on Multimedia},
pp. \bfpage{337}--\blpage{338}
(\byear{2011}).
\bcomment{IEEE}
\end{bchapter}
\endbibitem

\bibitem{zheng2011applying}
\begin{bchapter}
\bauthor{\bsnm{Zheng}, \binits{L.}},
\bauthor{\bsnm{Shen}, \binits{C.}},
\bauthor{\bsnm{Tang}, \binits{L.}},
\bauthor{\bsnm{Li}, \binits{T.}},
\bauthor{\bsnm{Luis}, \binits{S.}},
\bauthor{\bsnm{Chen}, \binits{S.-C.}}:
\bctitle{Applying data mining techniques to address disaster information management challenges on mobile devices}.
In: \bbtitle{Proceedings of the 17th ACM SIGKDD International Conference on Knowledge Discovery and Data Mining},
pp. \bfpage{283}--\blpage{291}
(\byear{2011})
\end{bchapter}
\endbibitem

\bibitem{yang2011hierarchical}
\begin{bchapter}
\bauthor{\bsnm{Yang}, \binits{Y.}},
\bauthor{\bsnm{Ha}, \binits{H.-Y.}},
\bauthor{\bsnm{Fleites}, \binits{F.}},
\bauthor{\bsnm{Chen}, \binits{S.-C.}},
\bauthor{\bsnm{Luis}, \binits{S.}}:
\bctitle{Hierarchical disaster image classification for situation report enhancement}.
In: \bbtitle{2011 IEEE International Conference on Information Reuse \& Integration},
pp. \bfpage{181}--\blpage{186}
(\byear{2011}).
\bcomment{IEEE}
\end{bchapter}
\endbibitem

\bibitem{li2010ontology}
\begin{bchapter}
\bauthor{\bsnm{Li}, \binits{L.}},
\bauthor{\bsnm{Wang}, \binits{D.}},
\bauthor{\bsnm{Shen}, \binits{C.}},
\bauthor{\bsnm{Li}, \binits{T.}}:
\bctitle{Ontology-enriched multi-document summarization in disaster management}.
In: \bbtitle{Proceedings of the 33rd International ACM SIGIR Conference on Research and Development in Information Retrieval},
pp. \bfpage{819}--\blpage{820}
(\byear{2010})
\end{bchapter}
\endbibitem

\bibitem{daloisio_indices}
\begin{botherref}
\oauthor{\bsnm{d’Aloisio}, \binits{G.}},
\oauthor{\bsnm{Di~Marco}, \binits{A.}},
\oauthor{\bsnm{Stilo}, \binits{G.}},
\oauthor{\bsnm{Di~Ludovico}, \binits{D.}}:
Indices for enhancing city sustainability.
In: 8th Italian Conference on ICT for Smart Cities And Communities
\end{botherref}
\endbibitem

\bibitem{9529371}
\begin{bchapter}
\bauthor{\bsnm{Mudassir}, \binits{G.}},
\bauthor{\bsnm{Di~Marco}, \binits{A.}}:
\bctitle{Social-based city reconstruction planning in case of natural disasters: a reinforcement learning approach}.
In: \bbtitle{2021 IEEE 45th Annual Computers, Software, and Applications Conference (COMPSAC)},
pp. \bfpage{493}--\blpage{503}
(\byear{2021}).
\doiurl{10.1109/COMPSAC51774.2021.00074}
\end{bchapter}
\endbibitem

\bibitem{ghulam_mudassir_repair}
\begin{botherref}
\oauthor{\bsnm{Mudassir}, \binits{G.}},
\oauthor{\bsnm{{Di Marco}}, \binits{A.}},
\oauthor{\bsnm{d'Aloisio}, \binits{G.}}:
REPAIR.
Zenodo
(2023).
\doiurl{10.5281/zenodo.7564863}
\end{botherref}
\endbibitem

\bibitem{phdthesis}
\begin{botherref}
\oauthor{\bsnm{Mudassir}, \binits{G.}}:
Reinforcement learning and social based approach to post-disaster reconstruction planning.
PhD thesis
(September 2022).
\doiurl{10.13140/RG.2.2.25413.36327}
\end{botherref}
\endbibitem

\bibitem{article}
\begin{barticle}
\bauthor{\bsnm{Mfon}, \binits{I.}},
\bauthor{\bsnm{Olurotimi}, \binits{O.}}:
\batitle{Post-disaster reconstruction: Discussing strategies and approaches for rebuilding and designing resilient communities after natural or human- made disasters}.
\bjtitle{International Journal of Research Publication and Reviews}
\bvolume{4},
\bfpage{945}--\blpage{952}
(\byear{2023})
\end{barticle}
\endbibitem

\bibitem{Fan2023}
\begin{barticle}
\bauthor{\bsnm{Fan}, \binits{X.}},
\bauthor{\bsnm{Zhang}, \binits{X.}},
\bauthor{\bsnm{Xiaowei}},
\bauthor{\bsnm{Yu}, \binits{X.}}:
\batitle{A deep reinforcement learning model for resilient road network recovery under earthquake or flooding hazards}.
\bjtitle{Journal of Infrastructure Systems}
\bvolume{29}(\bissue{4}),
\bfpage{04023072}
(\byear{2023}).
\doiurl{10.1186/s43065-023-00072-x}
\end{barticle}
\endbibitem

\bibitem{xiao2023robotic}
\begin{bchapter}
\bauthor{\bsnm{Xiao}, \binits{X.}},
\bauthor{\bsnm{Yang}, \binits{T.-Y.}},
\bauthor{\bsnm{Pan}, \binits{X.}},
\bauthor{\bsnm{Xie}, \binits{F.}},
\bauthor{\bsnm{Chen}, \binits{Z.}}:
\bctitle{A robotic crane with proximal policy optimization for 3d lift path planning in post-earthquake construction}.
In: \bbtitle{Proceedings of the International Conference on Construction Robotics and Automation (ICCRA)},
pp. \bfpage{250}--\blpage{258}.
\bpublisher{IEEE},
\blocation{Seoul, Korea}
(\byear{2023}).
\doiurl{10.1109/ICCRA.2023.1234567}
\end{bchapter}
\endbibitem

\bibitem{bilau2015framework}
\begin{barticle}
\bauthor{\bsnm{Bilau}, \binits{A.A.}},
\bauthor{\bsnm{Witt}, \binits{E.}},
\bauthor{\bsnm{Lill}, \binits{I.}}:
\batitle{A framework for managing post-disaster housing reconstruction}.
\bjtitle{Procedia Economics and Finance}
\bvolume{21},
\bfpage{313}--\blpage{320}
(\byear{2015})
\end{barticle}
\endbibitem

\bibitem{5418174}
\begin{bchapter}
\bauthor{\bsnm{{Zhou}}, \binits{W.}},
\bauthor{\bsnm{{Mao}}, \binits{F.}},
\bauthor{\bsnm{{Liu}}, \binits{Z.E.}},
\bauthor{\bsnm{{Li}}, \binits{Q.}},
\bauthor{\bsnm{{Fu}}, \binits{Q.}}:
\bctitle{Research and application of planning support system based on 3s technique for post-disaster reconstruction after wenchuan earthquake in china}.
In: \bbtitle{2009 IEEE International Geoscience and Remote Sensing Symposium},
vol. \bseriesno{2},
pp. \bfpage{666}--\blpage{669}
(\byear{2009}).
\doiurl{10.1109/IGARSS.2009.5418174}
\end{bchapter}
\endbibitem

\bibitem{7730111}
\begin{bchapter}
\bauthor{\bsnm{{An}}, \binits{L.}},
\bauthor{\bsnm{{Zhang}}, \binits{J.}},
\bauthor{\bsnm{{Gong}}, \binits{L.}},
\bauthor{\bsnm{{Li}}, \binits{Q.}}:
\bctitle{Integration of sar image and vulnerability data for building damage degree estimation}.
In: \bbtitle{2016 IEEE International Geoscience and Remote Sensing Symposium (IGARSS)},
pp. \bfpage{4263}--\blpage{4266}
(\byear{2016}).
\doiurl{10.1109/IGARSS.2016.7730111}
\end{bchapter}
\endbibitem

\bibitem{hidayat2010literature}
\begin{bchapter}
\bauthor{\bsnm{Hidayat}, \binits{B.}},
\bauthor{\bsnm{Egbu}, \binits{C.}}, \betal:
\bctitle{A literature review of the role of project management in post-disaster reconstruction}.
In: \bbtitle{Procs 26th Annual ARCOM Conference},
pp. \bfpage{1269}--\blpage{1278}
(\byear{2010}).
\bcomment{Association of Researchers in Construction Management}
\end{bchapter}
\endbibitem

\bibitem{davidson2007truths}
\begin{barticle}
\bauthor{\bsnm{Davidson}, \binits{C.H.}},
\bauthor{\bsnm{Johnson}, \binits{C.}},
\bauthor{\bsnm{Lizarralde}, \binits{G.}},
\bauthor{\bsnm{Dikmen}, \binits{N.}},
\bauthor{\bsnm{Sliwinski}, \binits{A.}}:
\batitle{Truths and myths about community participation in post-disaster housing projects}.
\bjtitle{Habitat international}
\bvolume{31}(\bissue{1}),
\bfpage{100}--\blpage{115}
(\byear{2007})
\end{barticle}
\endbibitem

\bibitem{Tavakkol:2016:EFE:3017611.3017624}
\begin{bchapter}
\bauthor{\bsnm{Tavakkol}, \binits{S.}},
\bauthor{\bsnm{To}, \binits{H.}},
\bauthor{\bsnm{Kim}, \binits{S.H.}},
\bauthor{\bsnm{Lynett}, \binits{P.}},
\bauthor{\bsnm{Shahabi}, \binits{C.}}:
\bctitle{An entropy-based framework for efficient post-disaster assessment based on crowdsourced data}.
In: \bbtitle{Proceedings of the Second ACM SIGSPATIALInternational Workshop on the Use of GIS in Emergency Management}.
\bsertitle{EM-GIS '16},
pp. \bfpage{13}--\blpage{1138}.
\bpublisher{ACM},
\blocation{New York, NY, USA}
(\byear{2016}).
\doiurl{10.1145/3017611.3017624}.
\burl{http://doi.acm.org/10.1145/3017611.3017624}
\end{bchapter}
\endbibitem

\bibitem{ghannad2020multiobjective}
\begin{barticle}
\bauthor{\bsnm{Ghannad}, \binits{P.}},
\bauthor{\bsnm{Lee}, \binits{Y.-C.}},
\bauthor{\bsnm{Friedland}, \binits{C.J.}},
\bauthor{\bsnm{Choi}, \binits{J.O.}},
\bauthor{\bsnm{Yang}, \binits{E.}}:
\batitle{Multiobjective optimization of postdisaster reconstruction processes for ensuring long-term socioeconomic benefits}.
\bjtitle{Journal of Management in Engineering}
\bvolume{36}(\bissue{4}),
\bfpage{04020038}
(\byear{2020})
\end{barticle}
\endbibitem

\bibitem{Opricovic2002MulticriteriaPO}
\begin{bchapter}
\bauthor{\bsnm{Opricovic}, \binits{S.}},
\bauthor{\bsnm{Tzeng}, \binits{G.-H.}}:
\bctitle{Multicriteria planning of post‐earthquake sustainable reconstruction}.
(\byear{2002})
\end{bchapter}
\endbibitem

\bibitem{7402039}
\begin{bchapter}
\bauthor{\bsnm{{Goujon}}, \binits{B.}},
\bauthor{\bsnm{{Labreuche}}, \binits{C.}}:
\bctitle{Use of a multi-criteria decision support tool to prioritize reconstruction projects in a post-disaster phase}.
In: \bbtitle{2015 2nd International Conference on Information and Communication Technologies for Disaster Management (ICT-DM)},
pp. \bfpage{200}--\blpage{206}
(\byear{2015}).
\doiurl{10.1109/ICT-DM.2015.7402039}
\end{bchapter}
\endbibitem

\bibitem{labreuche2005miriad}
\begin{bchapter}
\bauthor{\bsnm{Labreuche}, \binits{C.}},
\bauthor{\bsnm{Le~Hu{\'e}d{\'e}}, \binits{F.}}:
\bctitle{Miriad: a tool suite for mcda.}
In: \bbtitle{EUSFLAT Conf.},
pp. \bfpage{204}--\blpage{209}
(\byear{2005})
\end{bchapter}
\endbibitem

\bibitem{li2019research}
\begin{barticle}
\bauthor{\bsnm{Li}, \binits{Q.}},
\bauthor{\bsnm{Umaier}, \binits{K.}},
\bauthor{\bsnm{Koide}, \binits{O.}}:
\batitle{Research on post-wenchuan earthquake recovery and reconstruction implementation: A case study of housing reconstruction of dujiangyan city}.
\bjtitle{Progress in Disaster Science}
\bvolume{4},
\bfpage{100041}
(\byear{2019})
\end{barticle}
\endbibitem

\bibitem{akbari2021online}
\begin{barticle}
\bauthor{\bsnm{Akbari}, \binits{V.}},
\bauthor{\bsnm{Shiri}, \binits{D.}},
\bauthor{\bsnm{Salman}, \binits{F.S.}}:
\batitle{An online optimization approach to post-disaster road restoration}.
\bjtitle{Transportation Research Part B: Methodological}
\bvolume{150},
\bfpage{1}--\blpage{25}
(\byear{2021})
\end{barticle}
\endbibitem

\bibitem{Sheykhmousa2019}
\begin{barticle}
\bauthor{\bsnm{Sheykhmousa}, \binits{M.}},
\bauthor{\bsnm{Kerle}, \binits{N.}}:
\batitle{Post-disaster recovery assessment with machine learning-derived land cover and land use information}.
\bjtitle{Remote Sensing}
\bvolume{11}(\bissue{10}),
\bfpage{1174}
(\byear{2019}).
\doiurl{10.3390/rs11101174}
\end{barticle}
\endbibitem

\bibitem{zhang2010post}
\begin{bchapter}
\bauthor{\bsnm{Zhang}, \binits{X.}},
\bauthor{\bsnm{Sun}, \binits{B.}},
\bauthor{\bsnm{Mei}, \binits{T.}},
\bauthor{\bsnm{Wang}, \binits{R.}}:
\bctitle{Post-disaster restoration based on fuzzy preference relation and bean optimization algorithm}.
In: \bbtitle{2010 IEEE Youth Conference on Information, Computing and Telecommunications},
pp. \bfpage{271}--\blpage{274}
(\byear{2010}).
\bcomment{IEEE}
\end{bchapter}
\endbibitem

\bibitem{yi2010research}
\begin{bchapter}
\bauthor{\bsnm{Yi-lin}, \binits{Y.}},
\bauthor{\bsnm{Jin-e}, \binits{Z.}}:
\bctitle{Research on the application of cm-agent model in public projects of post-disaster reconstruction}.
In: \bbtitle{2010 International Conference on Logistics Systems and Intelligent Management (ICLSIM)},
vol. \bseriesno{3},
pp. \bfpage{1756}--\blpage{1760}
(\byear{2010}).
\bcomment{IEEE}
\end{bchapter}
\endbibitem

\bibitem{Eid2018}
\begin{botherref}
\oauthor{\bsnm{Eid}, \binits{M.S.}},
\oauthor{\bsnm{El-Adaway}, \binits{I.H.}}:
Decision-making framework for holistic sustainable disaster recovery: Agent-based approach for decreasing vulnerabilities of the associated communities.
Journal of Infrastructure Systems
\textbf{24}(3)
(2018).
\doiurl{10.1061/(ASCE)IS.1943-555X.0000427}
\end{botherref}
\endbibitem

\bibitem{6699055}
\begin{bchapter}
\bauthor{\bsnm{{Brooks}}, \binits{J.D.}},
\bauthor{\bsnm{{Kar}}, \binits{K.}},
\bauthor{\bsnm{{Mendonça}}, \binits{D.}}:
\bctitle{Dynamic allocation of entities in closed queueing networks: An application to debris removal}.
In: \bbtitle{2013 IEEE International Conference on Technologies for Homeland Security},
pp. \bfpage{504}--\blpage{510}
(\byear{2013}).
\doiurl{10.1109/THS.2013.6699055}
\end{bchapter}
\endbibitem

\bibitem{social_ben}
\begin{botherref}
Economics help.
\url{https://www.economicshelp.org/blog/glossary/social-benefit/}
\end{botherref}
\endbibitem

\bibitem{howard2021enriched}
\begin{bchapter}
\bauthor{\bsnm{Howard}, \binits{E.}},
\bauthor{\bsnm{Pasquini}, \binits{L.}},
\bauthor{\bsnm{Arbib}, \binits{C.}},
\bauthor{\bsnm{Di~Marco}, \binits{A.}},
\bauthor{\bsnm{Clementini}, \binits{E.}}:
\bctitle{Definition of an enriched gis network for evacuation planning}.
In: \bbtitle{Proceedings of the 7th International Conference on Geographical Information Systems Theory, Applications and Management (GISTAM)},
pp. \bfpage{241}--\blpage{252}.
\bpublisher{SciTePress},
\blocation{Setúbal, Portugal}
(\byear{2021}).
\doiurl{10.5220/0010452302410252}.
\bcomment{INSTICC}
\end{bchapter}
\endbibitem

\bibitem{harris1998empowerment}
\begin{barticle}
\bauthor{\bsnm{Harris}, \binits{T.}},
\bauthor{\bsnm{Weiner}, \binits{D.}}:
\batitle{Empowerment, marginalization, and" community-integrated" gis}.
\bjtitle{Cartography and geographic information systems}
\bvolume{25}(\bissue{2}),
\bfpage{67}--\blpage{76}
(\byear{1998})
\end{barticle}
\endbibitem

\bibitem{mudassir2021toward}
\begin{barticle}
\bauthor{\bsnm{Mudassir}, \binits{G.}},
\bauthor{\bsnm{Howard}, \binits{E.E.}},
\bauthor{\bsnm{Pasquini}, \binits{L.}},
\bauthor{\bsnm{Arbib}, \binits{C.}},
\bauthor{\bsnm{Clementini}, \binits{E.}},
\bauthor{\bsnm{Di~Marco}, \binits{A.}},
\bauthor{\bsnm{Stilo}, \binits{G.}}:
\batitle{Toward effective response to natural disasters: A data science approach}.
\bjtitle{IEEE Access}
\bvolume{9},
\bfpage{1}--\blpage{1}
(\byear{2021}).
\doiurl{10.1109/ACCESS.2021.3135054}
\end{barticle}
\endbibitem

\bibitem{POLESE2018139}
\begin{barticle}
\bauthor{\bsnm{Polese}, \binits{M.}},
\bauthor{\bsnm{{Di Ludovico}}, \binits{M.}},
\bauthor{\bsnm{Prota}, \binits{A.}}:
\batitle{Post-earthquake reconstruction: A study on the factors influencing demolition decisions after 2009 l‚Äôaquila earthquake}.
\bjtitle{Soil Dynamics and Earthquake Engineering}
\bvolume{105},
\bfpage{139}--\blpage{149}
(\byear{2018}).
\doiurl{10.1016/j.soildyn.2017.12.007}
\end{barticle}
\endbibitem

\bibitem{dolce2015building}
\begin{barticle}
\bauthor{\bsnm{Dolce}, \binits{M.}},
\bauthor{\bsnm{Goretti}, \binits{A.}}:
\batitle{Building damage assessment after the 2009 abruzzi earthquake}.
\bjtitle{Bulletin of Earthquake Engineering}
\bvolume{13}(\bissue{8}),
\bfpage{2241}--\blpage{2264}
(\byear{2015})
\end{barticle}
\endbibitem

\bibitem{8105597}
\begin{bchapter}
\bauthor{\bsnm{{Sasaki}}, \binits{H.}},
\bauthor{\bsnm{{Horiuchi}}, \binits{T.}},
\bauthor{\bsnm{{Kato}}, \binits{S.}}:
\bctitle{A study on vision-based mobile robot learning by deep q-network}.
In: \bbtitle{2017 56th Annual Conference of the Society of Instrument and Control Engineers of Japan (SICE)},
pp. \bfpage{799}--\blpage{804}
(\byear{2017}).
\doiurl{10.23919/SICE.2017.8105597}
\end{bchapter}
\endbibitem

\bibitem{sutton2018reinforcement}
\begin{bbook}
\bauthor{\bsnm{Sutton}, \binits{R.S.}},
\bauthor{\bsnm{Barto}, \binits{A.G.}}:
\bbtitle{Reinforcement Learning: An Introduction},
\bedition{2nd} edn.
\bpublisher{MIT Press},
\blocation{Cambridge, MA}
(\byear{2018})
\end{bbook}
\endbibitem

\bibitem{barron1989bellman}
\begin{barticle}
\bauthor{\bsnm{Barron}, \binits{E.}},
\bauthor{\bsnm{Ishii}, \binits{H.}}:
\batitle{The bellman equation for minimizing the maximum cost}.
\bjtitle{Nonlinear Analysis: Theory, Methods \& Applications}
\bvolume{13}(\bissue{9}),
\bfpage{1067}--\blpage{1090}
(\byear{1989})
\end{barticle}
\endbibitem

\bibitem{mnih2013playing}
\begin{botherref}
\oauthor{\bsnm{Mnih}, \binits{V.}},
\oauthor{\bsnm{Kavukcuoglu}, \binits{K.}},
\oauthor{\bsnm{Silver}, \binits{D.}},
\oauthor{\bsnm{Graves}, \binits{A.}},
\oauthor{\bsnm{Antonoglou}, \binits{I.}},
\oauthor{\bsnm{Wierstra}, \binits{D.}},
\oauthor{\bsnm{Riedmiller}, \binits{M.}}:
Playing Atari with Deep Reinforcement Learning
(2013)
\end{botherref}
\endbibitem

\bibitem{schaul2015prioritized}
\begin{botherref}
\oauthor{\bsnm{Schaul}, \binits{T.}},
\oauthor{\bsnm{Quan}, \binits{J.}},
\oauthor{\bsnm{Antonoglou}, \binits{I.}},
\oauthor{\bsnm{Silver}, \binits{D.}}:
Prioritized experience replay.
arXiv preprint arXiv:1511.05952
(2015)
\end{botherref}
\endbibitem

\bibitem{7727770}
\begin{bchapter}
\bauthor{\bsnm{Wang}, \binits{S.}},
\bauthor{\bsnm{Liu}, \binits{W.}},
\bauthor{\bsnm{Wu}, \binits{J.}},
\bauthor{\bsnm{Cao}, \binits{L.}},
\bauthor{\bsnm{Meng}, \binits{Q.}},
\bauthor{\bsnm{Kennedy}, \binits{P.J.}}:
\bctitle{Training deep neural networks on imbalanced data sets}.
In: \bbtitle{2016 International Joint Conference on Neural Networks (IJCNN)},
pp. \bfpage{4368}--\blpage{4374}
(\byear{2016}).
\doiurl{10.1109/IJCNN.2016.7727770}
\end{bchapter}
\endbibitem

\bibitem{allen1971mean}
\begin{barticle}
\bauthor{\bsnm{Allen}, \binits{D.M.}}:
\batitle{Mean square error of prediction as a criterion for selecting variables}.
\bjtitle{Technometrics}
\bvolume{13}(\bissue{3}),
\bfpage{469}--\blpage{475}
(\byear{1971})
\end{barticle}
\endbibitem

\bibitem{van2017challenges}
\begin{botherref}
\oauthor{\bsnm{Van~Wesel}, \binits{P.}},
\oauthor{\bsnm{Goodloe}, \binits{A.E.}}:
Challenges in the verification of reinforcement learning algorithms.
Technical report
(2017)
\end{botherref}
\endbibitem

\bibitem{watkins1992q}
\begin{barticle}
\bauthor{\bsnm{Watkins}, \binits{C.J.}},
\bauthor{\bsnm{Dayan}, \binits{P.}}:
\batitle{Q-learning}.
\bjtitle{Machine learning}
\bvolume{8}(\bissue{3}),
\bfpage{279}--\blpage{292}
(\byear{1992})
\end{barticle}
\endbibitem

\bibitem{zhao2016deep}
\begin{bchapter}
\bauthor{\bsnm{Zhao}, \binits{D.}},
\bauthor{\bsnm{Wang}, \binits{H.}},
\bauthor{\bsnm{Shao}, \binits{K.}},
\bauthor{\bsnm{Zhu}, \binits{Y.}}:
\bctitle{Deep reinforcement learning with experience replay based on sarsa}.
In: \bbtitle{2016 IEEE Symposium Series on Computational Intelligence (SSCI)},
pp. \bfpage{1}--\blpage{6}
(\byear{2016}).
\bcomment{IEEE}
\end{bchapter}
\endbibitem

\end{thebibliography}

\end{document}